\documentclass[prd,aps,reprint,groupedaddress,floatfix,flushbottom,nobibnotes,nofootinbib,showpacs,amsmath]{revtex4-1}
\usepackage{amssymb}
\usepackage{graphicx}
\usepackage{longtable}
\usepackage{hyperref}
\usepackage{braket}
\usepackage{mathbbol}
\begin{document}
\def\Regensburg{Institute for Theoretical Physics, University of Regensburg, 93040 Regensburg, Germany}
\title{Masses and decay constants of the $D_{s0}^*(2317)$ and $D_{s1}(2460)$ from $N_f=2$ lattice QCD close to the physical point}
\author{Gunnar S.~Bali}
\altaffiliation[Adjunct Faculty: ]{Tata Institute of Fundamental Research, Homi Bhabha Road, Mumbai 400005, India}
\affiliation{\Regensburg}
\author{Sara~Collins}
\email{sara.collins@ur.de}
\affiliation{\Regensburg}
\author{Antonio Cox}
\email{antonio.cox@ur.de}
\affiliation{\Regensburg}
\author{Andreas~Sch\"{a}fer}  
\affiliation{\Regensburg}
\collaboration{RQCD Collaboration}
\date{\today}
\begin{abstract}
We perform a high statistics study of the $J^{P}=0^{+}$ and $1^{+}$
charmed-strange mesons, $D_{s0}^*(2317)$ and $D_{s1}(2460)$,
respectively. The effects of the nearby $DK$ and $D^{*}K$ thresholds
are taken into account by employing the corresponding four quark
operators. Six ensembles with $N_f=2$ non-perturbatively ${\cal O}(a)$
improved clover Wilson sea quarks at $a=0.07$~fm are employed,
covering different spatial volumes and pion masses: linear lattice
extents $L/a=24,32,40,64$, equivalent to 1.7~fm to 4.5~fm, are
realised for $m_{\pi}=290$ MeV and $L/a=48,64$ or 3.4~fm and 4.5~fm
for an almost physical pion mass of $150$ MeV. Through a phase shift
analysis and the effective range approximation we determine the
scattering lengths, couplings to the thresholds and the infinite
volume masses.  Differences relative to the experimental values are
observed for these masses, however, this is likely to be due to
discretisation effects as spin-averaged quantities and splittings are
reasonably compatible with experiment. We also compute the weak decay
constants of the scalar and axialvector and find
$f_V^{0^+}=114(2)(0)(+5)(10)$~MeV and
$f_A^{1^+}=194(3)(4)(+5)(10)$~MeV, where the errors are due to
statistics, renormalisation, finite volume and lattice spacing
effects.
\end{abstract}
\maketitle

\section{Introduction}

In 2003 the BABAR Collaboration announced the observation of a
meson state in the inclusive $D_s^+\pi^0$ invariant mass
distribution~\cite{Aubert:2003fg}, compatible with a $J^{P}=0^{+}$
assignment, the $D_{s0}^*(2317)$. This discovery was confirmed soon
after by the CLEO and Belle
Collaborations~\cite{Besson:2003jp,Krokovny:2003zq}.  The newfound
state was the natural candidate to fill in the charm-strange $0^{+}$
$P$-wave level predicted by quark models.  However, while quark
models~\cite{Godfrey:1985xj,Godfrey:1986wj} and a number of early lattice
calculations~\cite{Lewis:2000sv,Hein:2000qu,Bali:2003jv,Dougall:2003hv} based on
quark-antiquark interpolators predicted the $0^{+}$ state to be a
broad resonance above the nearby $DK$ threshold, the experiments
observed a narrow state of mass $2317$ MeV, $40$ MeV below threshold.
The detection of another narrow state just below the $D^{*}K$
threshold, the
$D_{s1}(2460)$~\cite{Besson:2003cp,Abe:2003jk,Aubert:2003pe} with
$J^P=1^+$, presented a similar puzzle.

The strange-charm meson sector can be interpreted within heavy quark
effective
theory~\cite{Isgur:1991wq,Nowak:1992um,Bardeen:1993ae,Neubert:1993mb,Ebert:1994tv,Bardeen:2003kt}~(HQET).
At leading order in the inverse of the heavy quark mass, the states
are arranged in degenerate doublets corresponding to the strange quark
quantum numbers: $j^{P}=\frac{1}{2}^{-}$ for angular momentum $l=0$
and $j^P=\frac{1}{2}^{+}$ and $\frac{3}{2}^{+}$ for $l=1$ and so
on. Interactions beyond leading order, including with the heavy
(charm) quark spin, lift the degeneracies and cause mixing between
$j^P=\frac{1}{2}^+$ and $\frac{3}{2}^{+}$ states. The relevant quantum
numbers are then the total quark and antiquark spin, i.e. $J^P=0^-$,
$1^-$, for the $l=0$ doublet and $0^+$, $1^+$ and $1^+$, $2^+$ for
$l=1$. The doublets can be (loosely) identified with the observed
$\left(D_{s},D_{s}^{*}\right)$,
$\left(D_{s0}^{*}(2317),D_{s1}\left(2460\right)\right)$ and
$\left(D_{s1}\left(2536\right),D_{s2}^{*}(2573)\right)$ mesons,
respectively. Nevertheless, the surprisingly low masses of the
$D_{s0}^{*}(2317)$ and $D_{s1}\left(2460\right)$ mesons have led to a
number of more exotic interpretations, for example, as
tetraquarks~\cite{Barnes:2003dj,Terasaki:2003qa,Chen:2004dy},
molecules~\cite{Cheng:2003kg,Browder:2003fk} or conventional
charm-strange mesons with coupled channel
effects~\cite{vanBeveren:2003kd}.  A recent comprehensive review of
the experimental status and theoretical understanding of these states
can be found in Ref.~\cite{Chen:2016spr}.

Subsequent lattice
studies~\cite{Mohler:2011ke,Namekawa:2011wt,Moir:2013ub}, utilising
quark-antiquark interpolators and, most recently, including chiral and
continuum extrapolations~\cite{Cichy:2016bci} also overestimate the
mass of the $D_{s0}^*(2317)$. A similarly conventional analysis by
some of us found consistency with the $0^+$ and $1^+$ $D_s$
experimental masses in Ref.~\cite{Bali:2015lka}, however, there were a
number of systematic uncertainties that could not be quantified. The
possible influence of the nearby threshold needs to be taken into
account by incorporating four-quark $DK$ interpolators and performing
a finite volume analysis utilising L\"{u}scher's
formalism~\cite{Luscher:1990ux} for the
unequal mass case~\cite{Davoudi:2011md,Fu:2011xz,Leskovec:2012gb}.
The first work in this direction was
performed by Liu and collaborators who computed the scattering lengths
for the $D\overline{K}$ system for which there are no computationally
challenging disconnected diagrams~\cite{Liu:2012zya}. Predictions were
made for the $DK$ channel via SU(3) flavour symmetry. Following this,
Mohler et al.~\cite{Mohler:2013rwa} and Lang et
al.~\cite{Lang:2014yfa} studied the $D_{s0}^*(2317)$ and
$D_{s1}(2460)$ mesons directly, including coupling with the threshold,
and found their masses to be compatible with experiment for an
ensemble with $m_\pi=156$~MeV, at a fairly coarse lattice spacing of
$a=0.09$~fm and a small spatial lattice extent of
$L=2.9$~fm~($Lm_\pi=2.29$). The effective range approximation was
assumed in order to extract infinite volume results. Notably, the
masses of these states were found to be overestimated if the $DK$
interpolators were omitted.

Clearly, a number of improvements can be made on this pioneering study
working, for example, at a finer lattice spacing and
exploring the dependence on the spatial volume. The former is important
since discretisation effects can be substantial for observables
involving charm quarks while the latter is needed as contributions
which are exponentially suppressed in $Lm_\pi$~(that are ignored in
the L\"uscher formalism) may not be small for
$Lm_\pi=2.29$. Furthermore, the range of validity of the effective
range approximation needs to be tested.  

In this work we present a high statistics analysis at $a=0.07$~fm for
two pion masses, $m_\pi=290$ and 150~MeV, utilising multiple spatial
volumes, with $L$ in the range of $1.7$ to 4.5~fm realising values for
$Lm_\pi$ between $2.7$ to $6.7$. Near to physical pion masses are
required as the $0^+$ and $1^+$ charm-strange states are sensitive to
the position of the threshold and one needs to reproduce the physical
case. By employing $N_f=2$ dynamical fermions, effects arising from strange sea
quarks are omitted with the expectation that the valence strange quark
provides the dominant contribution. Furthermore, we treat the
$D_{s0}^*(2317)$ and $D_{s1}(2460)$ as stable and ignore their
(strong) decays to $D_s\pi$ and $D_s^*\pi$ and $D_s\pi\pi$,
respectively. This is reasonable, given that the first two decays are
isospin-violating (and in our simulation isospin is exact) and the
third has a very small width. Effects of the higher lying $D_s\eta$
and $D^*_s\eta$ thresholds are also neglected.

\begin{table*}
\caption{Details of the ensembles used for this analysis. $Lm_\pi$ is computed using the
infinite volume pion mass determined in Ref.~\cite{Bali:2014nma}.
}\label{tab_1}
\begin{ruledtabular}
\begin{tabular}{cccccccccc}
$\kappa_{l}$ & $a$ [fm]& $V$ & $am_\pi$ & $m_\pi$ [MeV] &
$Lm_\pi$ & $m_K$~[MeV] & $m_D$~[MeV] & $m_{D^{*}}$~[MeV] & $N_{\rm conf}$
\\ \hline 
0.13632 & 0.071 & $24^3\times 48$ & 0.1112(9) & 306.9(2.5) & 2.67 & $540(2)$ & $1907(3)$ & $2038(5)$ & $2222$ \\ 
& 0.071 & $32^3\times 64$ & 0.10675(52) & 294.6(1.4) & 3.42 & $528(1)$ & $1902(3)$ & $2030(5)$ & $1453$ \\ 
& 0.071 & $40^3\times 64$ & 0.10465(38) & 288.8(1.1) & 4.19 & $527(1)$ & $1901(2)$ & $2030(4)$ & $2000$ \\ 
& 0.071 & $64^3\times 64$ & 0.10487(24) & 289.5(0.7) & 6.70 & $526(1)$ & $1898(1)$ & $2030(2)$ & $1463$ \\ \hline 
0.13640 & 0.071 & $48^3\times 64$ & 0.05786(55) & 159.7(1.5) & 2.78 & $500(1)$ & $1880(2)$ & $2007(3)$ & $2501$ \\ 
& 0.071 & $64^3\times 64$ & 0.05425(49) & 149.7(1.4) & 3.49 & $497(1)$ & $1877(1)$ & $1996(3)$ & $1591$ 
\end{tabular}
\end{ruledtabular}
\end{table*}

So far, most lattice studies have focused on computing the particle
masses and the couplings of the states to the two meson channels.  In
this work, we also determine the weak decay constants, 
i.e. the overlap of the (local) weak current operator with the
physical state, for $J^P=0^+$ and the lower $1^+$ meson. The decay
constants have not yet been directly determined in experiment,
however, some information can be extracted from non-leptonic $B$
decays to $D^{(*)}D_{sJ}^{(*)}$. Within the factorisation
approximation, invoking the heavy quark
limit~\cite{Beneke:2000ry,Luo:2001mc}, ratios of the corresponding
branching fractions give $f_{D_{s0}^*(2317)}\sim f_{Ds}/3$, while for
the axialvector channel $f_{D_{s1}(2460)}\sim 2f_{D^*_s}/3$, see, for
example, the analyses of
Refs.~\cite{Datta:2003re,Hwang:2004kga,Cheng:2006dm}.  These results,
however, are at odds with heavy quark symmetry studies which find
$f_{D_{s1}(2460)}\sim
f_{D_{s0}^*(2317)}$~\cite{Colangelo:1999ny,LeYaouanc:2001ma,Colangelo:2002dg}.
The decay constants have also been computed, for example, within quark
models~\cite{LeYaouanc:2001ma,Cheng:2003sm,Hsieh:2003xj,Verma:2011yw,Segovia:2012yh}
and QCD sum rules~\cite{Colangelo:2005hv,Wang:2015mxa} with results
covering a wide range, $f_{D_{s0}^*(2317)}=70-440$~MeV and
$f_{D_{s1}(2460)}=117-410$~MeV.

The paper is organised as follows. Details of the lattice set-up are
given in Section~\ref{latsetup}. The construction of the quark line
diagrams required for extracting the energy levels and matrix elements
for the states of interest are discussed in
Section~\ref{correlator}. The procedure for extracting the phase
shifts, the couplings to the two meson channels and the masses from
the finite volume levels is well established and we only provide a
brief overview of the theoretical background in
Section~\ref{sec:theo}. We extract the infinite volume information
employing two methods: L\"uscher's formalism~\cite{Luscher:1990ux,Davoudi:2011md,Fu:2011xz,Leskovec:2012gb} as
well as the chiral unitary approach~\cite{Oller:1997ti,Oller:1997ng},
which also allows us to determine the so-called potential of the
scattering particles. Our results on the phase shifts, scattering
lengths, potentials, spectrum and decay constants are presented in
Section~\ref{res}, before we conclude in Section~\ref{conc}.

\section{Lattice set-up}
\label{latsetup}

\begin{figure}
\centerline{
\includegraphics[width=.48\textwidth,clip=]{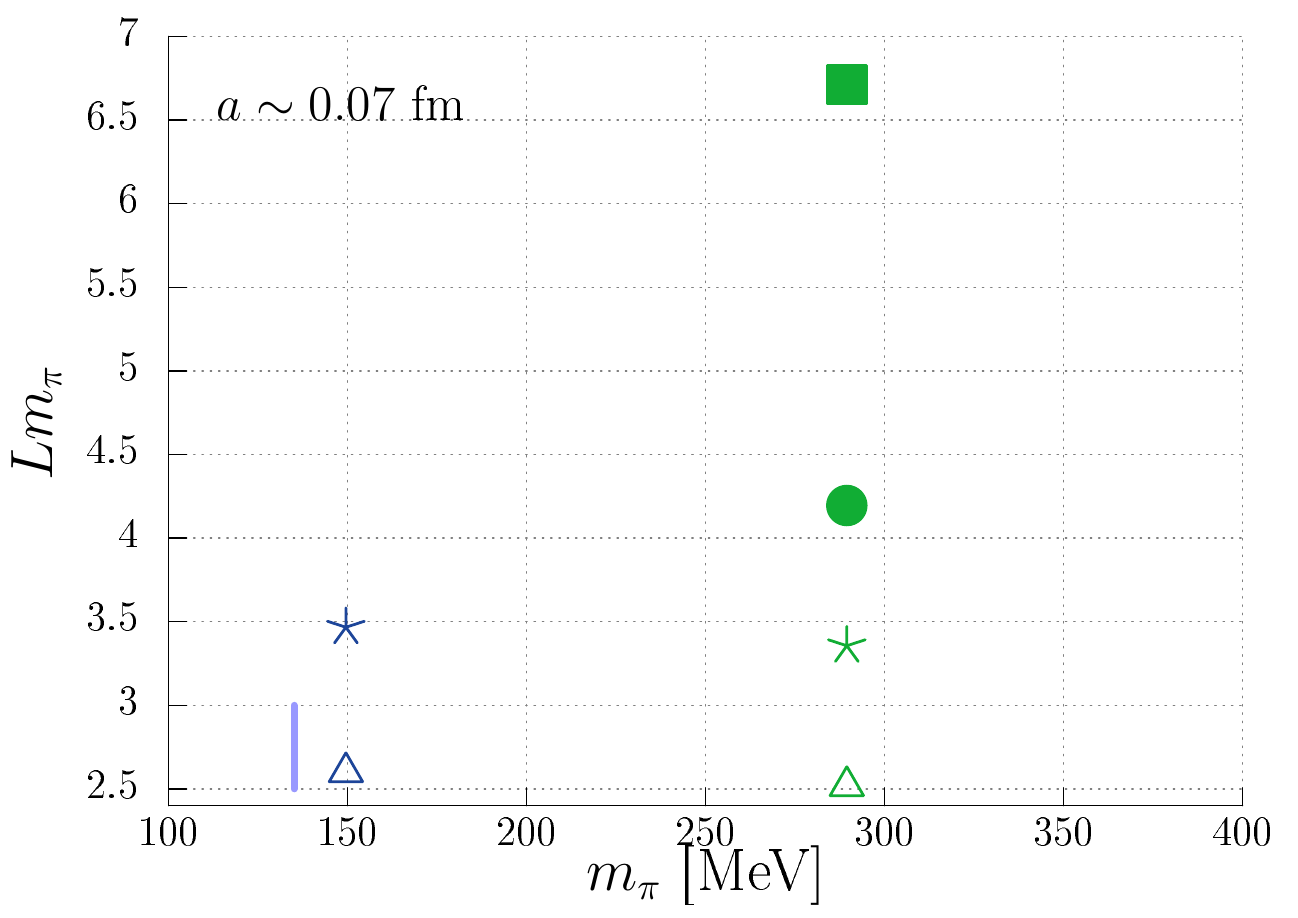}
}
\caption{Overview of the ensembles employed in our analysis  in terms of the pion
mass and the spatial extent $L$~(in units of $m_\pi$). The vertical line indicates
the physical pion mass.}
\label{fig_ens}
\end{figure}

In order to study the volume dependence of the lowest lying energy
levels, various spatial volumes are realised at two pion masses,
$m_{\pi} \sim290$ MeV with $L/a=24,32,40,64$ and $m_{\pi} \sim150$ MeV
with $L/a=48,64$, where $L$ denotes the linear extent.  The ensembles
were generated by the RQCD and QCDSF collaborations and are composed
of $N_{f}=2$ non-perturbatively improved clover fermions at a single
lattice spacing $a=0.071$ fm~\cite{Bali:2014nma}~(determined via the
Sommer scale $r_0$~\cite{Sommer:1993ce}).  Details of the ensembles
are given in Table~\ref{tab_1} and Fig.~\ref{fig_ens}.  The strange
and charm quarks are partially quenched in our analysis and their
masses are fixed by reproducing~(to within 1$\%$) the combination
$\sqrt{2m_{K}^{2}-m_{\pi}^{2}}=685.8$~MeV employing the electrically
neutral, isospin-averaged estimates from the FLAG
review~\cite{Aoki:2016frl}~(see the discussion below) and the
experimental value of the spin-averaged $1S$ charmonium mass,
$m_{1S}=3068.5$~MeV, respectively. When computing the latter we omit
disconnected quark line diagrams and mixing with other flavour
singlets. The effect of this omission is likely to be only a few MeV
in the $1S$ charmonium mass~(see, for example, the studies in
Refs.~\cite{Levkova:2010ft,Bali:2011rd}) and does not lead to a
significant uncertainty in our results for the $D_s$ spectrum.

As mentioned previously, reproducing the physical $DK$ and $D^*K$
thresholds is important for studying the $0^+$ and $1^+$ states,
respectively. In order to compare our lattice values for these
thresholds and other levels with experiment, however, corrections are
required as we are working in the isospin limit and electromagnetic
effects are absent. We choose to adjust the experimental results
rather than correcting the lattice values.  For the kaon we take the
FLAG review~\cite{Aoki:2016frl} value of $m_K^{\rm QCD}=494.2(3)$~MeV for
the physical mass in QCD. For the $D^{(*)}$ mesons we define the
electrically neutral isospin symmetric mass as,
\begin{equation}
m^{\rm QCD}_{D^{(*)}}=\frac{1}{2}\left(m_{D^{(*)0}}+m_{D^{(*)\pm}}-\delta m_{D^{(*)}}^{\rm QED}\right).
\end{equation}
The electromagnetic mass contributions, $\delta m_D^{\rm QED}=2.3(2)$~MeV
and $\delta m_{D^*}^{\rm QED}=0.8(2)$~MeV were estimated in
Ref.~\cite{Goity:2007fu} in the heavy quark limit including $1/m_Q$
terms.  To be conservative we double the size of these QED
errors. Combining these values with the experimental masses gives
$m^{\rm QCD}_D=1866.1(2)$~MeV and $m^{\rm QCD}_{D^*}=2008.2(2)$~MeV. For the
$D_s^{(*)}$ mesons the electromagnetic mass contribution is assumed to
be of the same size as for the $D$ mesons with,
\begin{equation}
m^{\rm QCD}_{D_s^{(*)}}=m_{D_s^{(*)}}-\delta m_{D^{(*)}}^{\rm QED},
\end{equation}
giving $m_{D_s}^{\rm QCD}=1966.0(4)$~MeV and
$m_{D_s}^{\rm QCD}=2111.3(6)$~MeV.  No estimates have been made of $\delta
m^{\rm QED}$ for the positive parity charm-strange mesons and in this case
we add an additional error of 2~MeV to the experimental masses to
indicate the likely size of this uncertainty.  So, for example, we
quote for the $0^+$ mass, $m_{0^+}=2317.7(0.6)(2.0)$~MeV, where the
first error is experimental, while for the splitting with the
threshold we give $m_K+m_D-m_{0^+}=42.6(0.7)(2.0)$~MeV, with the first
error due to the QCD estimate of $m_K+m_D$.
Turning to the lattice data in Table~\ref{tab_1} for the $m_\pi=150$~MeV,
$L=64a$ ensemble, the kaon mass is compatible with the FLAG estimate,
while the $D$~($D^*$) meson mass is slightly above~(below) the
QCD value. This leads to the $DK$ and $D^*K$ thresholds being missed
by only $+14$ and $-9$~MeV, respectively.

Leading order discretisation effects are of
$\mathcal{O}(a^{2})$ and, as the charm quark mass in
lattice units is not small ($am_{c}\sim0.5$), lattice spacing effects
can be significant. Fine structure splittings are expected to be
particularly sensitive to such effects as they are dominated by
momentum scales close to $m_{c}$ for heavy-light systems.  This is
illustrated by our results for the $D$ and $D_s$ $1S$ hyperfine
splittings, $m_{D^{*}}-m_{D}=119(3)$ MeV and
$m_{D_{s}^{*}}-m_{D_{s}}=118(1)$ MeV, from the largest $m_\pi=150$~MeV
ensemble, which are approximately 23~MeV and 27~MeV below the
corrected experimental values, respectively. 
In contrast, spin-averaged splittings which have typical
energy scales that are much smaller than the inverse lattice
spacing~(of the order of $\overline{\Lambda}\sim 0.5$~GeV for heavy-light
systems which is much less than $a^{-1}=2.76$~GeV), are less affected
as will be demonstrated in Section~\ref{res}.

We perform a high statistics study utilising 1450 to 2200
configurations for each ensemble, see Table~\ref{tab_1}. Careful
consideration of auto-correlations is required and these were taken
into account by binning over measurements~(one per configuration) to a
level consistent with at least four times the integrated
auto-correlation time.
 
Finite volume effects on hadron masses and decay constants fall off
exponentially with $Lm_{\pi}$ and empirically $Lm_{\pi}>4$ has been
found to be sufficient for such effects to be suppressed in most
observables. In L\"{u}scher's formalism smaller volumes are beneficial
for obtaining infinite volume information, however, the exponentially
suppressed finite volume terms are neglected and $Lm_{\pi}$ cannot be
too small. This will be discussed in Section~\ref{res}; for our ensembles 
$Lm_{\pi}$ ranges from 2.67 to 6.71.

\section{Correlator matrix}
\label{correlator}

Two distinct sectors corresponding to $J^P=0^+$ and $1^+$ are
considered in this work.  In the first case, the lowest energy level
is expected to coincide with the bound state $D_{s0}^*(2317)$, followed
by a $DK$ scattering state somewhat above.  Analogously, in the
second case we expect to find the $D_{s1}(2460)$, followed by a
$D^*K$ scattering state as well as the $D_{s1}(2536)$.

In order to extract these levels a variational analysis is
performed~\cite{Michael:1985ne,Luscher:1990ck}. Choosing a set of
quark-antiquark and two meson interpolators $O_{i}$ which have an overlap
$Z_{kj}=\braket{k|O_{j}^{\dagger}|0}$ with the physical states of
interest, $\ket{k}$, a correlator matrix is constructed,
\begin{eqnarray} 
C_{ij}\left(t\right)=\braket{0|O_{i}\left(t\right)O_{j}^{\dagger}|0}=\sum_{k}Z_{ik}^{\dagger}Z_{kj}e^{-E_{k}t} .\label{eq:17}
\end{eqnarray}
Note that the interpolators are projected onto zero momentum.
By solving the generalised eigenvalue equation 
\begin{eqnarray} 
C\left(t\right)v^{\left(k\right)}(t,t_0)=\lambda^{\left(k\right)}\left(t,t_{0}\right)C\left(t_{0}\right)v^{\left(k\right)}(t,t_0)\label{eq:18}
\end{eqnarray} 
for eigenvalues $\lambda^{\left(k\right)}(t,t_0)$ and eigenvectors
$v^{\left(k\right)}(t,t_0)$ for $t>t_0$, $t_{0}$ being a reference time,
the energy levels are obtained from the exponential decay of the
eigenvalues
\begin{eqnarray}
\lambda^{\left(k\right)}\left(t,t_{0}\right)=e^{-E_{k}\left(t-t_{0}\right)}\left(1+\mathcal{O}\left(e^{-\Delta E_{k}t}\right)\right),\label{eq:19} 
\end{eqnarray}  
where $\Delta E_{k}$ is the difference between $E_{k}$ and the first
energy level outside of the rank of the basis considered for $t<2t_0$
and $t-t_0$ constant~\cite{Blossier:2009kd}.  Clearly, the basis of
operators must be large enough in order to resolve the number of
levels of interest, and in general, due to the contamination from
higher states one needs a basis of at least $n+1$ operators in order
to reliably extract $n$ states.

The choice of operators is also important, especially for the
charm-strange systems of interest here where the lowest two energy
levels are very close to each other~(in particular for the larger
spatial volumes): a basis of operators with poor overlap with the
physical states will not separate the energy levels within the finite
(Euclidean) time extent of the lattice.  This is precisely the problem
when forming a basis of only $\bar{q}q$ interpolators, which leads to
the overestimation of the mass of both the lowest $0^+$ and $1^+$
$D_{s}$ states as illustrated in
Refs.~\cite{Mohler:2013rwa,Lang:2014yfa} and demonstrated again in
Section~\ref{res:energies}.

\begin{table}
\caption{Interpolators used in the analysis. 
}\label{tab_2}
\begin{ruledtabular}
\begin{tabular}{cl}$J^{P}$ &  Two-quark operators \\
\hline 
$0^{+}$ &  $O_{D_{s}}=\overline{s}\mathbb{1}c,\quad O_{D'_{s}}=\overline{s}\gamma_{t}c$ \\
$1^{+}$ & $O_{D_{s}}=\overline{s}\gamma_{i}\gamma_{5}c,\quad O_{D'_{s}}=\overline{s}\gamma_{t}\gamma_{i}\gamma_{5}c$ \\\hline
$J^{P}$ &  Four-quark operators \\\hline 
$0^{+}$ & $O_{DK}=\left(\overline{u}\gamma_{5}c\right)\left(\overline{s}\gamma_{5}u\right)+\left(\overline{d}\gamma_{5}c\right)\left(\overline{s}\gamma_{5}d\right)$\\
$1^{+}$ & $O_{DK}=\left(\overline{u}\gamma_{i}c\right)\left(\overline{s}\gamma_{5}u\right)+\left(\overline{d}\gamma_{i}c\right)\left(\overline{s}\gamma_{5}d\right)$
\end{tabular}
\end{ruledtabular}
\end{table}

Our interpolator basis includes both $\bar{q}q$ and four quark
operators and the correlator
matrix has the general form
\begin{eqnarray} C\left(t\right)=\left(\begin{array}{cc}
\braket{O_{D_{s}}\left(t\right)O_{D_{s}}^{\dagger}\left(0\right)} & \braket{O_{D_{s}}\left(t\right)O_{DK}^{\dagger}\left(0\right)}\\
\braket{O_{DK}\left(t\right)O_{D_{s}}^{\dagger}\left(0\right)} & \braket{O_{DK}\left(t\right)O_{DK}^{\dagger}\left(0\right)}
\end{array}\right),\,\,
\label{eq:20} 
\end{eqnarray} 
where ``$D_s$'' and ``$DK$'' denote the two and four quark cases,
respectively.  Several two quark interpolators are employed with
multiple smearing levels~(see Table~\ref{tab_2} and the discussion below),
such that the entries in Eq.~(\ref{eq:20}) represent sub-matrices.
The correlators are projected onto zero-momentum and for the two meson
interpolators, both the particles are at rest. The omission of
operators of the form $D(\boldsymbol{p})K(-\boldsymbol{p})$ for momentum $\boldsymbol{p}$
is discussed in Section~\ref{res:energies}. We remark that operators
with derivatives were also included in the analysis but the resulting
correlation functions were later discarded as they were too noisy.

The operators given in Table~\ref{tab_2} for the scalar and axialvector
channels fall in the $A_1$ and $T_1$ irreducible representations of the
lattice cubic group, respectively. These representations create a
tower of states which, in the continuum limit, correspond to
$J=0,4,6,\ldots$ and $J=1,3,4,\ldots$, and include ground (single
particle) states, radial excitations and multi-particle levels. As we
are only interested in the lowest $J$ in each case and the other
states lie much higher in the spectrum, there is very little
ambiguity in the spin identification of the energy levels we extract
and throughout this work we only refer to the lowest continuum spin created.

\begin{figure}
\centerline{
\includegraphics[width=.48\textwidth,clip=]{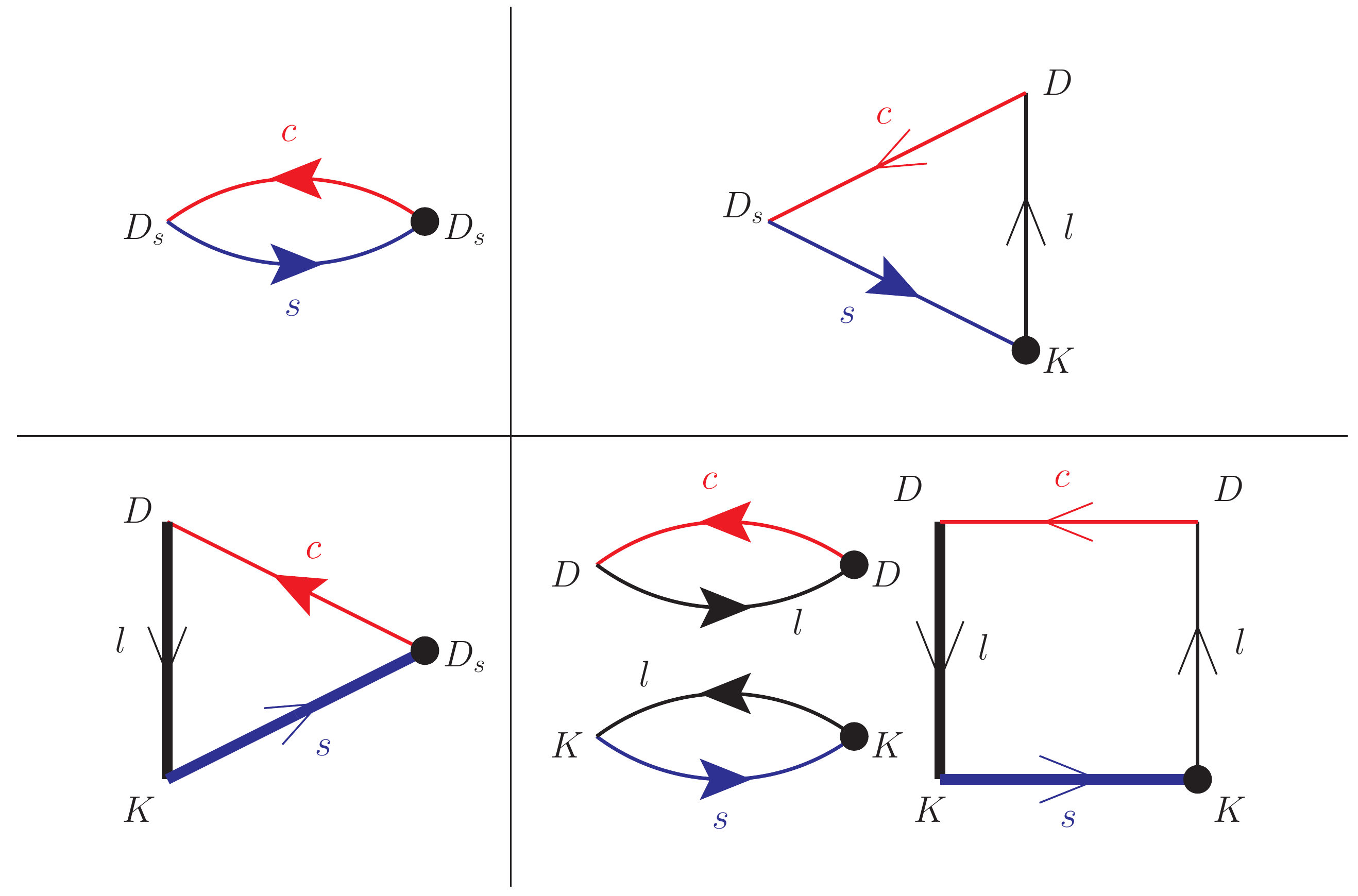}
}
\caption{The quark line diagrams computed on the lattice. The charm,
  strange and light quarks are indicated by red, blue and black lines,
  respectively.  Stochastic propagators are represented by lines with
  filled arrows and sequential stochastic propagators by two
  successive lines of the same width with open arrows. Time
  propagation is from right to left. The black dots indicate the
  stochastic source position. Note that the triangular diagrams are
  accompanied by a factor of 2 in Eq.~(\ref{eq:20}) and the $DK\to DK$
  diagrams are accompanied by a factor of 4 and 2 for the box and the
  product of $D$ and $K$ two-point functions, respectively, due to the
  summation over the light quark flavour, see Table~\ref{tab_2}. }
\label{fig_wick}
\end{figure}

The Wick contractions arising from Eq.~(\ref{eq:20}) are shown in
Fig.~\ref{fig_wick}. These quark line diagrams are evaluated using
spin and colour diluted complex $\mathbb{Z}_2$ stochastic sources with
the one-end trick~\cite{Foster:1998vw,McNeile:2002fh}, following
Refs.~\cite{Aoki:2007rd,Aoki:2011yj,Bali:2015gji}.
Evaluation of the $DK\to DK$ box diagram requires two sequential
propagators involving a combination of light and charm~($lc$) quarks
and strange and light quarks~($sl$), represented by the thin and thick
lines with open arrows in the bottom right of Fig.~\ref{fig_wick},
respectively. These sequential propagators are recycled in the
determination of the triangular diagrams that are averaged to improve
the signal. The other propagators required~(see the lines with filled
arrows in Fig.~\ref{fig_wick}) are similarly recycled where possible. 

The $sl$ sequential combination is the most computationally expensive
due to the need to realise the sequential source on every sink
timeslice $t$~(cf. Eq.~(\ref{eq:20})). For this reason we restrict
$t/a\in\left[5,19\right]$, a region chosen such that the excited state
contributions to the resulting correlation functions are not large and
the statistical noise is still under control.  This restriction
affects the box diagram and the lower left triangular diagram in
Fig.~\ref{fig_wick}. The remaining diagrams are evaluated for all
timeslices and the (anti-) periodic boundary conditions in the
temporal direction of length $T$ enable averaging over the time
regions $0<t<T/2$ and $T/2<t<T$.

Gauge noise was found to dominate the correlator matrix and only the
minimum number of stochastic sources was employed per
configuration. This corresponds to 12$\times$2, where the first factor
is due to spin-colour dilution and the second one arises because two
independent stochastic sources are required for the $DK\to DK$ diagram
involving the product of the $D$ and $K$ two-point functions. Spin
dilution is required in order to study both the $0^+$ and $1^+$ states
efficiently with the one-end trick. Colour dilution does not provide
any reduction in the stochastic noise for fixed computational cost,
however, implementing this within our code turned out to be
convenient.

\begin{figure}
\centerline{
\includegraphics[width=.48\textwidth,clip=]{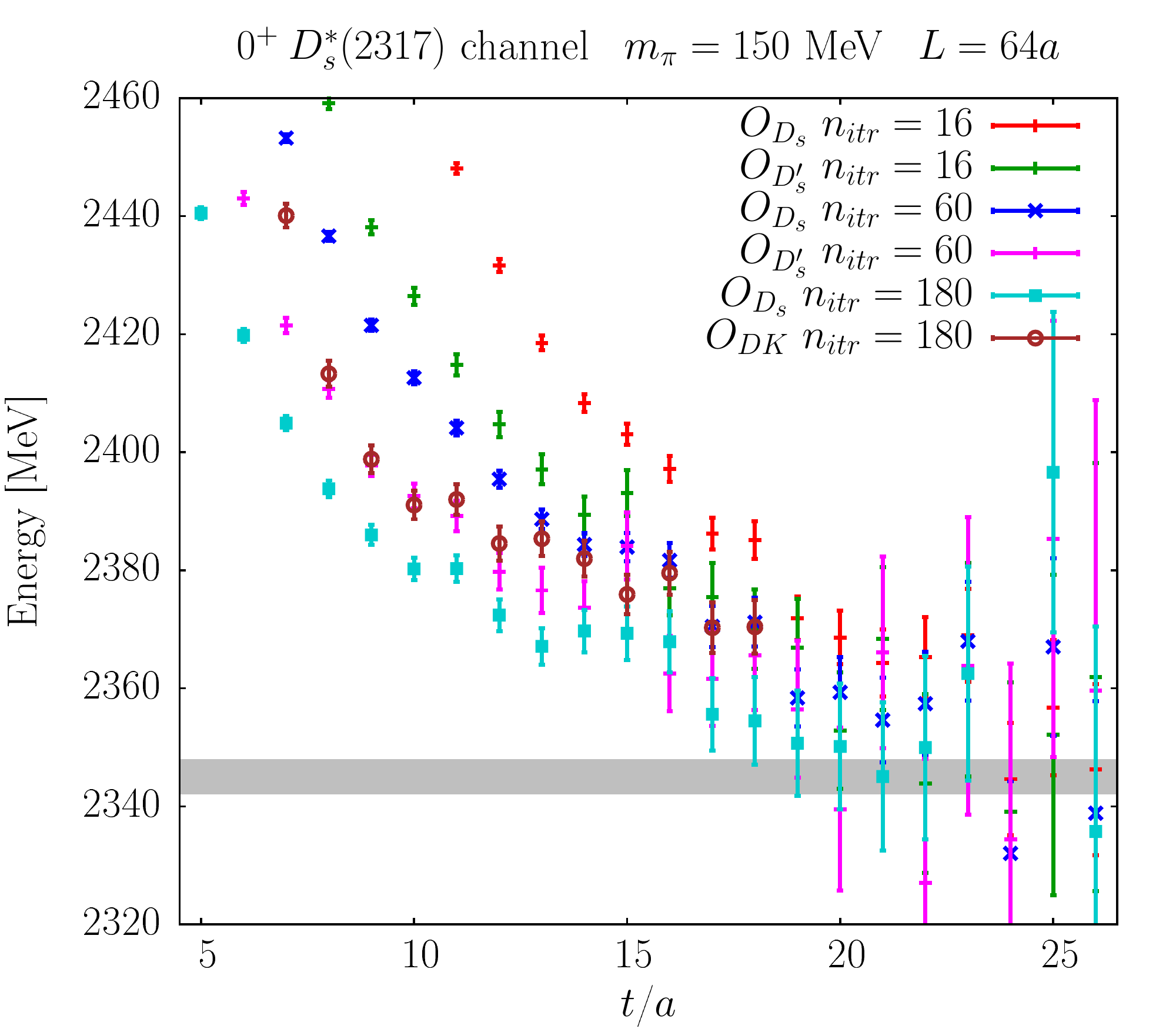}
}
\caption{The effective masses for the diagonal elements of the
  $6\times 6$ correlator matrix for the $0^+$ channel for
  $m_\pi=150$~MeV and $L/a=64$. The operator basis for the matrix
  consists of three operators of type $O_{D_{s}}$~(see
  Table~\ref{tab_2}) with different levels of Wuppertal smearing
  iterations, two of type $O_{D'_{s}}$ and one $O_{DK}$
  interpolator. The grey band indicates the ground state energy
  extracted by solving the generalised eigenvalue problem
  Eqs.~(\ref{eq:18}) and~(\ref{eq:19}), see Section~\ref{res} for
  details.}
\label{fig_effm_corrs}
\end{figure}

In order to ensure that for both the scalar and axialvector meson
sectors we can resolve at least the lowest three states, we construct
the $D_s$ and $D'_s$ operators~(see Table~\ref{tab_2}) with multiple
spatial extents and the $DK$ operators with a single spatial extent.
Wuppertal smearing~\cite{Gusken:1989ad} with 3 dimensionally APE
smoothed spatial links~\cite{Falcioni:1984ei,Albanese:1987ds} was
applied with the number of Wuppertal iterations~($n_{\rm itr}$) equal
to 16, 60 and $180$ for $O_{D_{s}}$ interpolators shared between quark
and antiquark, $n_{\rm itr}=16,60$ for $O_{D'_{s}}$ and $n_{\rm
  itr}=180$ for $O_{DK}$ operators.  These choices are illustrated for
the $0^+$ state in Fig.~\ref{fig_effm_corrs}, which displays the
effective masses\footnote{See Eq.~(\ref{eq:23}) for the definition of
  the effective mass.} of the diagonal components of the correlator
matrix. As expected, increasing $n_{\rm itr}$ significantly boosts the
overlap with the lowest state, at the cost of an increase in the noise
at larger times. Similar behaviour is observed for the $1^+$. The
determination of the lowest energy levels from the correlator matrix
via the variational approach is discussed in
Section~\ref{res:energies}, along with the impact of the operator
basis chosen. We also extract the decay constants of the
$D_{s0}^*(2317)$ and $D_{s1}(2460)$, as described in
Section~\ref{res:decay}. For this purpose we compute the diagrams in
the upper row of Fig.~\ref{fig_wick} with smeared source
interpolators and local $O_{D_s}$ and $O_{D'_s}$ sink operators.

In total a $6\times 6$ correlator matrix was realised at a
computational cost of 14 charm quark, 3 strange quark and $N_T+3$
light quark inversions for each spin and colour component of the
stochastic propagator~(i.e. times 12 for the full cost) per
configuration. We remark that in order to minimise the number of
inversions, the smearing for each operator was split unevenly between
the quark and antiquarks.  The number of timeslices, $N_T=15$, for
the light quark is due to the chosen range of the sink time mentioned
above.  The cost of these light quark inversions, equivalent to
$N_T+3$ point-to-all propagators, represents the main overhead
compared to a conventional analysis involving only quark-antiquark
operators. For the restricted basis of operators considered here, the
stochastic one end trick method we employed is substantially cheaper
than the distillation technique used in
Refs.~\cite{Mohler:2013rwa,Lang:2014yfa} and enabled much larger
lattice volumes to be realised. However, the latter approach becomes more
attractive when considering a wider range of the meson spectrum
involving several thresholds, see, for example,
Refs.~\cite{Moir:2013ub,Prelovsek:2014swa,Padmanath:2015jea,Moir:2016srx}.

\section{Theoretical background}
\label{sec:theo}
In the following we briefly outline how energy levels measured on a
finite lattice volume can be used to extract infinite volume
information via a parametrisation of the $T$-matrix.  Two approaches
are considered. The first is based on L\"uscher's formalism and the
effective range approximation, the second on a determination of the
potential of the scattering particles in the chiral unitary approach.

\subsection{L\"{u}scher's method and the effective range approximation}
\label{finitevolume}

For two relativistic particles with masses $m_{1}$ and $m_{2}$,
scattering elastically in infinite volume, the $s$-wave $T$-matrix in
the centre of momentum frame can be expressed as
\begin{equation}
T\left(s\right)=\frac{-8\pi\sqrt{s}}{p\cot\delta\left(p\right)-ip}, \label{eq:12}
\end{equation}
where $\sqrt{s}=E$ is the centre of momentum energy and
$p$ is the modulus of the momentum of each particle,
\begin{equation}
p^{2}=\frac{\left(s-\left(m_{1}+m_{2}\right)^{2}\right)\left(s-\left(m_{1}-m_{2}\right)^{2}\right)}{4s}. \label{eq:35}
\end{equation}
$\delta\left(p\right)$ is the $s$-wave phase shift and
$p\cot\delta\left(p\right)$ is a real function of $p^{2}$ which can be
expanded around the threshold $p^{2}=0$:
\begin{equation}
p\cot\delta\left(p\right)=\frac{1}{a_{0}}+\frac{1}{2}r_{0}p^{2}+\mathcal{O}\left(p^{4}\right).\label{eq:3} 
\end{equation}
The parameters $a_{0}$ and $r_{0}$ are the scattering
length  and the effective range, respectively, which, up
to $\mathcal{O}\left(p^{4}\right)$, describe the low-energy scattering
of the particles.

Above threshold, the $T$-matrix shows a unitarity cut which represents
the continuous spectrum. Here, unitarity dictates that the imaginary
part is given by,
$\mbox{Im}\,T^{-1}\left(s\right)=\frac{p}{8\pi\sqrt{s}}$.
Below threshold, $p=i|p|$ is imaginary and $T$ is real. If a bound
state is present at $s=s_{B}\equiv m_{B}^{2} $ or $p=p_{B}$, it will
appear as a pole of $T$ on the real axis:
\begin{equation}
p_{B}\cot\delta\left(p_{B}\right)=ip_{B}\equiv-|p_{B}|\label{eq:13}.
\end{equation}
In the vicinity of the pole the $T$-matrix takes the form
\begin{equation}
T\left(s\right)\sim \frac{g^{2}}{s-s_{B}},  \label{eq:36}
\end{equation}
so that the coupling $g$ can be obtained through
\begin{equation}
g^{2}=\lim_{s\rightarrow s_{B}}T\left(s\right)\left(s-s_{B}\right)=\lim_{s\rightarrow s_{B}}\frac{-8\pi\sqrt{s}\left(s-s_{B}\right)}{p\cot\delta\left(p\right)-ip}. \label{eq:30}
\end{equation}

At finite spatial volume, $L^{3}$, the energy levels and momenta are
discretised and the cut of the $T$-matrix is replaced by poles at
discrete values $s=s_{n}$:
\begin{eqnarray} 
\sqrt{s_{n}}=E_{n} & = & \sqrt{m_{1}^{2}+p_{n}^{2}}+\sqrt{m_{2}^{2}+p_{n}^{2}} \label{eq:1} \\ 
  & = & \sqrt{m_{1}^{2}+k_{n}^{2}}+\sqrt{m_{2}^{2}+k_{n}^{2}}+\Delta E_{n}, \label{eq:5}  
\end{eqnarray}
where $\boldsymbol{k_{n}}=\frac{2\pi}{L}\boldsymbol{n}$, $\boldsymbol{p_{n}}=\frac{2\pi}{L}\boldsymbol{q_{n}}$, $\boldsymbol{n}\in\mathbb{Z}^{3}$ and  $n=\sqrt{|\boldsymbol{n}|^{2}}=\sqrt{0},\sqrt{1},\sqrt{2},...$, while $\boldsymbol{q_n}$ are real valued vectors.
The asymptotic two particle states in the infinite volume formalism
are no longer free once placed in a finite box as the probability
for them to be within the interaction range is finite.
As $L$ increases, the interaction term $\Delta E_{n}$ tends to zero,
$\boldsymbol{q_{n}}\rightarrow\boldsymbol{n}$ and
$p^{2}_{n}\rightarrow k^{2}_{n}$.  The position of the bound state
pole $s_{B}$ is shifted to $s_{\widetilde{B}}$ at finite volume. We allow
the index $n$ to assume an additional value $n=\widetilde{B}$ so that in
Eqs.~(\ref{eq:1}) and (\ref{eq:5}) $E_{\widetilde{B}}=m_{\widetilde{B}}$,
$p_{\widetilde{B}}$ (imaginary) and $\Delta E_{\widetilde{B}}$ ($<0$)
represent, respectively, the mass, binding momentum and binding energy
of the bound state at finite volume. As $L\rightarrow\infty$, these
quantities will tend to their infinite volume values $m_{B}$, $p_{B}$
and $\Delta E_{B}$.

L{\"u}scher's equation~\cite{Luscher:1990ux} (and its analytical
continuation below threshold) relates the finite volume energy levels
to the (infinite volume) partial wave phase shift
$\delta\left(p\right)$. For $p=p_{n}$,
\begin{equation}
p\cot\delta\left(p\right)=\frac{2}{L\sqrt{\pi}}\mathcal{Z}_{00}\left(1;\frac{L^{2}}{4\pi^{2}}p^{2}\right),\label{eq:2}
\end{equation}
for $s$-wave scattering, where $\mathcal{Z}_{00}$ is the (analytic
continuation of the) generalised zeta-function.  The latter has a
simple exact expansion below threshold~\cite{Sasaki:2006jn}, so that
\begin{eqnarray} 
p\cot\delta\left(p\right) &=& ip+\frac{1}{L}\sum_{n=1}^{\infty}\frac{\theta_{n}}{\sqrt{n}}e^{-\sqrt{n}|p|L} \label{eq:14} \\
 & = & ip+\frac{1}{L}\left(6e^{-|p|L}+\frac{12}{\sqrt{2}}e^{-\sqrt{2}|p|L}+...\right), \nonumber
 \end{eqnarray} 
where $\theta_{n}$ is the theta series of a simple cubic lattice.  It
is clear that as $L$ increases the summation term approaches zero and
$p$ approaches the infinite volume binding momentum $p_{B}$ defined by
Eq.~(\ref{eq:13}). In principle, mixing with higher partial waves
needs to be considered when determining the phase shift. However,
these contributions are suppressed and for the energy range of
interest in this study, it is reasonable to neglect them.

Covering energies (through varying the lattice extent $L$) that are
below and above threshold, we compute $p\cot\delta\left(p\right)$ from
Eq.~(\ref{eq:2}) and perform the simple linear fit consistent with the effective
range approximation Eq.~(\ref{eq:3}) to determine $a_{0}$ and
$r_{0}$. Then the bound state condition Eq.~(\ref{eq:13}), which becomes
\begin{equation} 
\frac{1}{a_{0}}-\frac{1}{2}r_{0}|p_{B}|^{2}=-|p_{B}|, \label{eq:16}
\end{equation}  
will provide the infinite volume binding momentum $p_{B}$ and thus the
bound state mass $m_{B}$, using Eq.~(\ref{eq:1}).  Finally, the coupling
can be evaluated within the same approximation by expanding the
denominator of Eq.~(\ref{eq:30}) around $s_{B}\equiv m_{B}^{2}$ and
making use of Eq.~(\ref{eq:13}), to arrive at
\begin{equation} 
g^{2}=\frac{64\pi m_{B}p_{B}}{\left(1-r_{0}p_{B}\right)\left(1-\left(\frac{m_{1}^{2}-m_{2}^{2}}{m_{B}^{2}}\right)^{2}\right)}. \label{eq:31}
\end{equation}  

\subsection{Chiral unitary approach}

Within the chiral unitary approach the $s$-wave $T$-matrix is
expressed in terms of a (real) ``potential'' $V\left(s\right)$ for the
scattering particles,
\begin{equation}
T\left(s\right)=\frac{1}{V^{-1}\left(s\right)-G\left(s\right)}, \label{eq:32}
\end{equation}
and a loop function $G\left(s\right)$ of two meson propagators,
\begin{eqnarray}
G(s) & =
&\int_{|\boldsymbol{k}|<\Lambda}\frac{d^{3}k}{\left(2\pi\right)^{3}}I\left(s,\boldsymbol{k}\right),\label{eq:gs}
\end{eqnarray}
with
\begin{eqnarray}
I\left(s,\boldsymbol{k}\right)&=&\frac{1}{2\omega_{1}\left(\boldsymbol{k}\right)\omega_{2}\left(\boldsymbol{k}\right)}\frac{\omega_{1}\left(\boldsymbol{k}\right)+\omega_{2}\left(\boldsymbol{k}\right)}{s-\left(\omega_{1}\left(\boldsymbol{k}\right)+\omega_{2}\left(\boldsymbol{k}\right)\right)^{2}}\label{eq:41}
\end{eqnarray}
and
$\omega_{1/2}\left(\boldsymbol{k}\right)=\sqrt{m_{1/2}^{2}+\boldsymbol{k}^{2}}$.
The integral is divergent and can be regularised by imposing a cut-off
$\Lambda$ on the magnitude of $\boldsymbol{k}$. Alternatively, one can
perform dimensional regularisation and introduce a subtraction
constant, $\alpha(\mu)$, for a renormalisation scale $\mu$:
\begin{eqnarray}
G\left(s\right)&=&\frac{1}{16\pi^{2}}\left[\alpha\left(\mu\right)+\log\frac{m_{1}m_{2}}{\mu^{2}}+\right.\nonumber\\
 & & \hspace{2.5cm}\left.\frac{\delta
    m}{2s}\log\frac{m_{2}^{2}}{m_{1}^{2}}+\frac{p}{\sqrt{s}}l\left(s\right)\right] \label{eq:45}
\end{eqnarray}
and
\begin{eqnarray}
l\left(s\right)&=&+\log\left(2\sqrt{s}p+s+\delta
m\right)+\log\left(2\sqrt{s}p+s-\delta m\right) \nonumber
\\ &&-\log\left(2\sqrt{s}p-s+\delta
m\right)-\log\left(2\sqrt{s}p-s-\delta m\right),\nonumber\\&& \label{eq:37}
\end{eqnarray}
where $\delta m=m_{2}^{2}-m_{1}^{2}$ and $p$ is given by
Eq.~(\ref{eq:35}). 

With knowledge of the potential, the bound state mass, as a pole in
the $T$-matrix, can be obtained by imposing the condition
\begin{equation}
V\left(s_{B}\right)G\left(s_{B}\right)=1, \label{eq:44}
\end{equation} 
while in the vicinity of the pole one can combine the parametrisation
of Eq.~(\ref{eq:32}) with Eq.~(\ref{eq:36}) to derive the sum rule
\begin{equation}
\underbrace{g^{2}\frac{\partial V^{-1}}{\partial s}}_{Z}+\underbrace{g^{2}\left(-\frac{\partial G}{\partial s}\right)}_{1-Z}=1. \label{eq:33}
\end{equation}
We remark that in weakly coupled quantum mechanics the potential $V$
can be interpreted as a perturbation to a hypothetical,
non-interacting Hamiltonian $H_0$. Then $Z$ is the probability of the
bound state to correspond to the one-particle sector of $H_0$ while
$1-Z$ represents the probability that it is made up of more than one
free particle, e.g., the $D$ and the $K$. This is known as Weinberg's
compositeness condition~\cite{Weinberg:1965zz}. For detailed
discussions of the interpretation of this quantity within the present
context see, for example,
Refs.~\cite{Baru:2003qq,Gamermann:2009uq,Hyodo:2013nka}. However, it
is not clear how meaningful this notion is for a strongly
interacting quantum field theory. The nature of resonances in elastic
scattering with a nearby $s$-wave threshold
was earlier discussed in Refs.~\cite{Morgan:1992ge,Morgan:1993td}.

Note that the bound state mass, coupling and
``compositeness'' are independent of the choice of subtraction constant in
Eq.~(\ref{eq:45})~(or equivalently $\Lambda$ in Eq.~(\ref{eq:gs}))
since a change in $\alpha(\mu)$ is compensated for by a change in the potential such
that physical quantities remain unaffected.

Expressions for the (scalar) potential for $K$ and $D$ meson scattering can
be derived within heavy meson chiral perturbation
theory~\cite{Kolomeitsev:2003ac,Hofmann:2003je,Guo:2006fu,Gamermann:2006nm,Guo:2008gp,Guo:2009ct,Cleven:2010aw,Yao:2015qia,Guo:2015dha}~(HMChPT). At
leading order~\cite{Kolomeitsev:2003ac},
\begin{equation}
V\left(s\right)=\frac{1}{4F_{\pi}^{2}}\left[-3s+\frac{\left(m_{D}^{2}-m_{K}^{2}\right)^{2}}{s}+2\left(m_{D}^{2}+m_{K}^{2}\right)\right], \label{eq:38}
\end{equation}
where $F_{\pi}$ is the pion decay constant with the normalisation
corresponding to the experimental value of 92~MeV. However, the
potential can also be extracted using the energy spectrum determined
on the lattice.  Neglecting finite volume effects on the potential
that are exponentially suppressed, the $T$-matrix for a spatial extent $L$ reads
\begin{equation}
\widetilde{T}\left(s,L\right)=\frac{1}{V^{-1}\left(s\right)-\widetilde{G}\left(s,L\right)}. \label{eq:34}
\end{equation}
The finite volume loop function is normally expressed as the sum of
the infinite volume function~(given by Eq.~(\ref{eq:45})) and a
correction term $\Delta G\left(s,L\right)$,
\begin{eqnarray}
\widetilde{G}\left(s,L\right)&=&G\left(s\right)+\Delta
G\left(s,L\right),\label{eq:39} 
\end{eqnarray}
where,
\begin{eqnarray}
\Delta
G\left(s,L\right)&&=\nonumber\\ &&\hspace{-0.5cm}\lim_{\Lambda\rightarrow\infty}\left(\frac{1}{L^{3}}\sum_{\boldsymbol{k}}^{|\boldsymbol{k}|<\Lambda}-\int_{|\boldsymbol{k}|<\Lambda}\frac{d^{3}k}{\left(2\pi\right)^{3}}\right)I\left(s,\boldsymbol{k}\right).\label{eq:40}
\end{eqnarray}
The discrete sum is over the lattice momenta
$\boldsymbol{k}=\frac{2\pi}{L}\boldsymbol{n},\quad\boldsymbol{n}\in\mathbb{Z}^{3}$.
The lattice energy levels (squared), $s_{n}=s_{n}\left(L\right)$ in
Eq.~(\ref{eq:1}), correspond to poles of $\widetilde{T}$. Thus, the bound
state condition
\begin{equation}
V^{-1}\left(s_{n}\right)=\widetilde{G}\left(s_{n},L\right) \label{eq:42}
\end{equation}
allows us to probe the potential by evaluating
$\widetilde{G}\left(s_{n},L\right)$ for each $s_n\left(L\right)$.
Fitting the potential with a
modelling function~(see Section~\ref{res:pot}), the bound state mass
can be accessed by imposing Eq.~(\ref{eq:44}) and the coupling and
compositeness via Eq.~(\ref{eq:33}).

\begin{figure*}
\centerline{
\includegraphics[width=.9\textwidth,clip=]{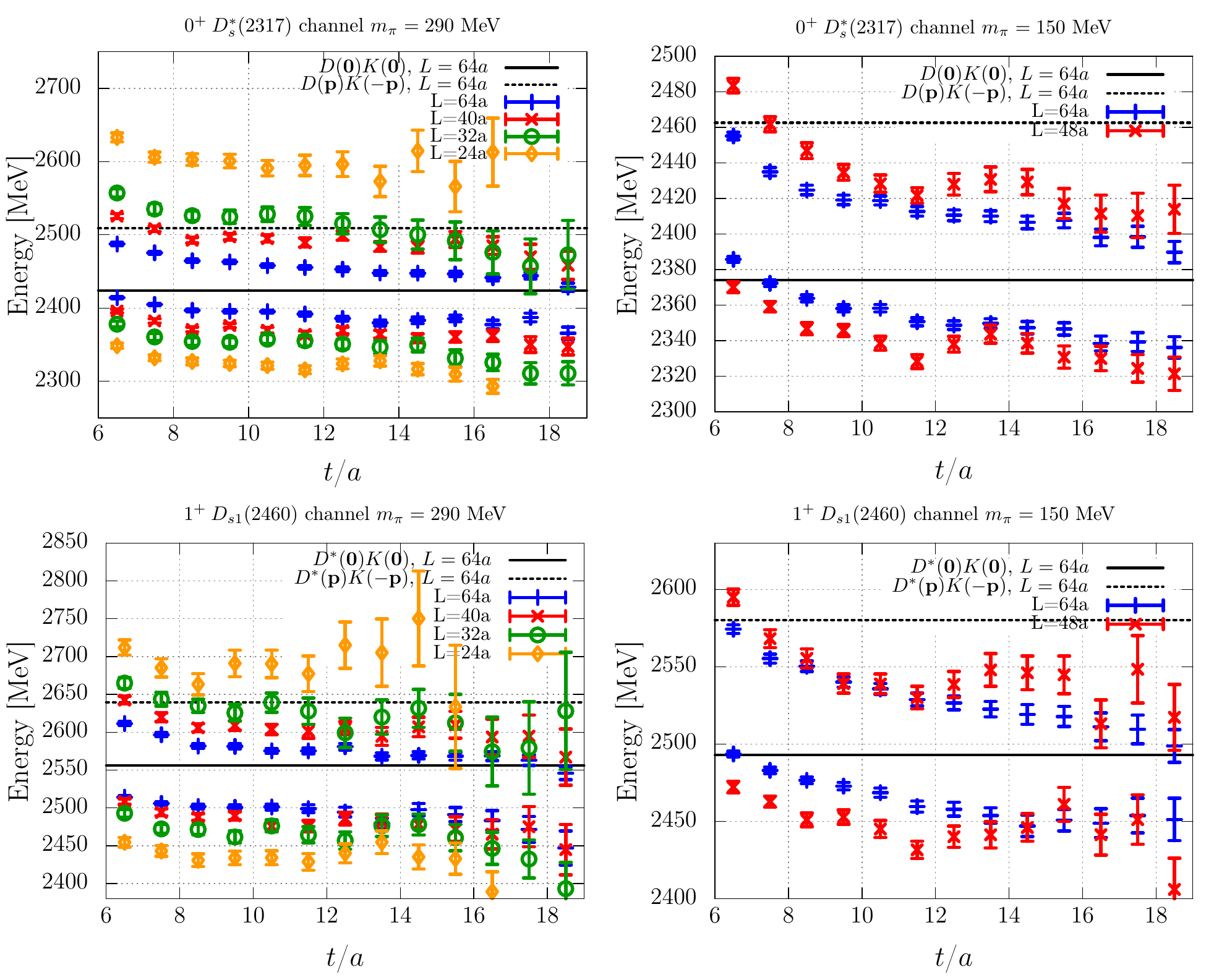}
}
\caption{The effective masses of the lowest two eigenvalues for the
  $0^+$~(top) and $1^+$~(bottom) sectors on ensembles with
  $m_\pi=290$~MeV~(left) and 150~MeV~(right).  The horizontal lines
  represent the lowest two free scattering states determined for the
  largest spatial volume at each pion mass, where for the second level
  corresponding to $D(\boldsymbol{p})K(-\boldsymbol{p})$, the spatial momentum
  $|\boldsymbol{p}|=2\pi/L$. A $4\times 4$ correlator matrix is employed in
  all the cases, consisting of the $O_{D^{(\prime)}_s}$ operators with three
  different smearing levels and the $O_{DK}$ operator with one
  smearing level~(see Section~\ref{correlator}). The energies of the third
  eigenvalues lie much higher. }
\label{fig:all_effm}
\end{figure*}

The infinite volume $T$-matrix can also be reconstructed:
\begin{equation}
T\left(s_{n}\right)=\frac{1}{\Delta G\left(s_{n},L\right)}. \label{eq:43}
\end{equation}
This is independent of the regulator used. Note that when extracting
the phase shift using Eq.~(\ref{eq:12}) an explicit form for the
potential does not have to be introduced. Indeed, as shown
in Ref.~\cite{Doring:2011vk}, this is a more general approach than
L\"{u}scher's, as small volume contributions are kept. However, in this
work, we find these additional contributions to be negligible.

\section{Results}
\label{res}

The matrix of correlators in Eq.~(\ref{eq:20}) is constructed for each
ensemble and the variational method applied. The extraction of the
(finite volume) spectra from the resulting eigenvalues is presented in
the next subsection. The phase shifts and infinite volume information,
including the masses and couplings, derived from the spectra via
L\"uscher's formalism are presented in Section~\ref{res:phase},
followed by a complementary analysis via the chiral unitary approach
in Section~\ref{res:pot}. Our results for the low lying $D_s$ spectrum
are given in Section~\ref{res:spec}.  In addition, we determine the
scalar and vector decay constants of the $D_{s0}^*(2317)$ and the
axialvector and tensor decay constants of the $D_{s1}^*(2460)$ in
Section~\ref{res:decay}.

\subsection{Energies}
\label{res:energies}

For each channel of interest the operator basis for constructing the
correlator matrix in Eq.~(\ref{eq:20}) is varied in order to determine
the influence of each interpolator on the energy spectrum and to realise
the best signals possible. Considering the $0^+$ channel first, a basis
of four operators consisting of $O_{D_s}$ with all three smearing levels
and $O_{DK}$ with a single smearing level~(see
Section~\ref{correlator} and Table~\ref{tab_2}) proved sufficient for
extracting the lowest two energies corresponding to the bound state
and the scattering state, as demonstrated below. The quality of the
signal achieved is illustrated in Fig.~\ref{fig:all_effm}, which
displays the effective masses,
\begin{equation} 
E_{n}\left(t+a/2,t_{0}\right)=\log\frac{\lambda_{n}\left(t,t_{0}\right)}{\lambda_{n}\left(t+a,t_{0}\right)},
 \label{eq:23}
\end{equation}  
for the two levels on all ensembles in the time range
$t/a\in\left[6,19\right]$ where $t>t_0$ and $t_0$ is set to
$5a$. Utilising higher values of $t_0$ gave consistent results. As
discussed in Section~\ref{correlator}, the range of $t$ is 
smaller than the lattice temporal extent as the computational cost in
terms of the number of light quark inversions for some elements of the
correlator matrix is roughly proportional
to the number of sink timeslices.

\begin{figure}
\centerline{
\includegraphics[width=.5\textwidth,clip=]{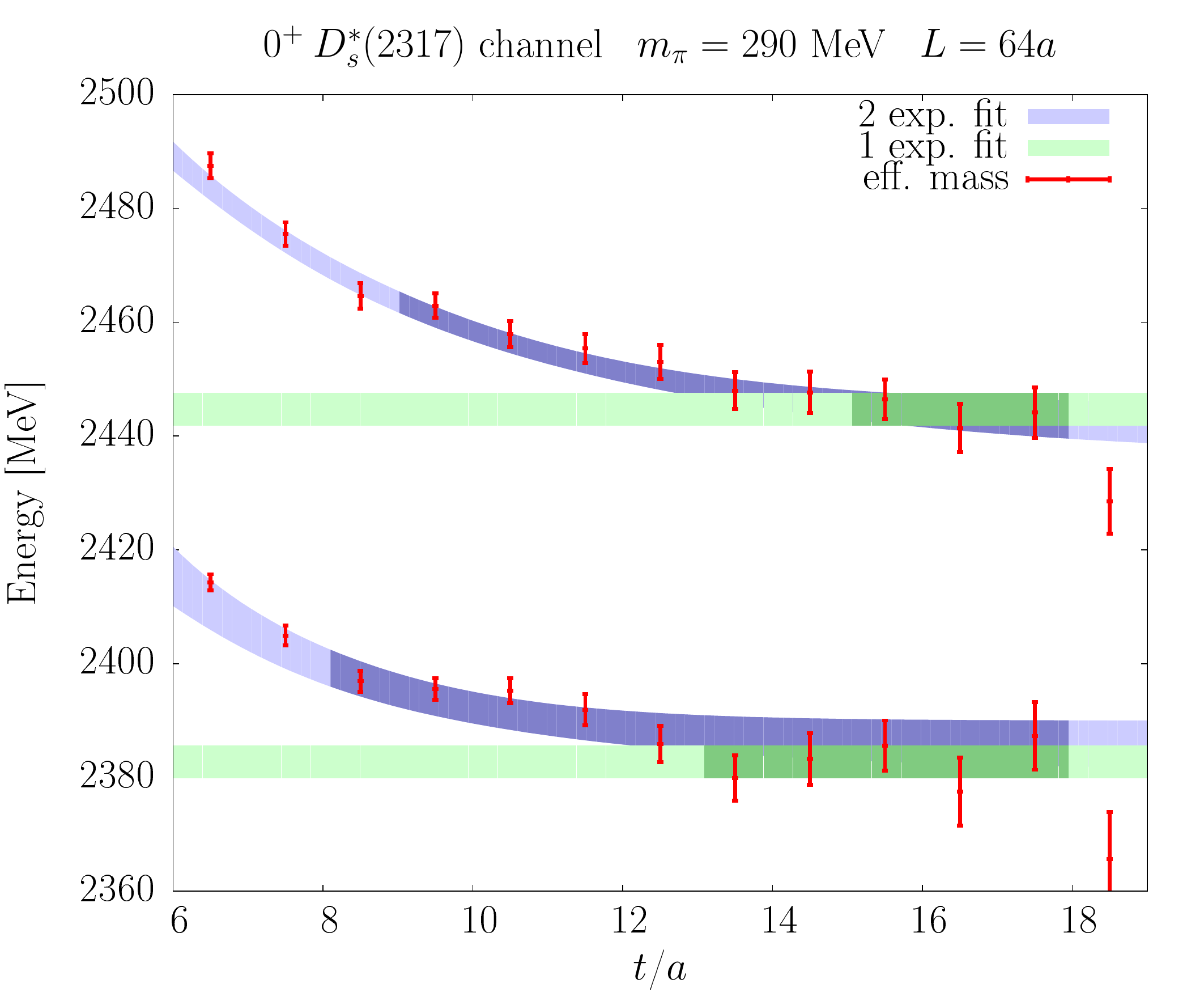}
}
\caption{The effective masses of the lowest two eigenvalues in the
  $0^{+}$ channel compared to the results from one and two exponential
  fits indicated by the green and blue bands, respectively, for the
  $m_\pi=290$~MeV, $L/a=64$ ensemble. The fitting ranges in each
  case are marked by the darker colours. The eigenvalues are generated
  from a $4\times 4$ correlator matrix as in Fig.~\ref{fig_effm_corrs}.}
\label{fig:expvs2exp}
\end{figure}

Figure~\ref{fig:all_effm} shows that unwanted contributions to the
eigenvalues from other (higher) states die away around timeslices
12--14 corresponding to the physical distances 0.8--1.0~fm. As the
spatial volume is increased the energy of the lowest state increases
and the next level decreases, tending towards the non-interacting
threshold. This behaviour is compatible with that of a bound
state~(the $D_{s0}^*(2317)$) that couples to the $DK$ threshold and a
scattering state.  The final results for the energies are extracted by
fitting the eigenvalues within a chosen time window. The end point for
the fit~($t_{\rm max}$) needs to be fixed with care due to the short physical
time extent of the lattices, corresponding to 3.4~fm for $L=24a$ and
4.5~fm for $L>24a$.  For (anti) periodic boundary conditions in the
temporal direction, there are additional contributions to the spectral
decomposition of $C_{ij}(t)$ in Eq.~(\ref{eq:17}). These include terms
arising from backward propagation in time of the form $Z_{ki}
Z_{kj}^\dagger e^{-E_k(T-t)}$, which can be neglected for $t<T/2$ in
our analysis due to the size of $E_k$ and $T$. However, there are also
so-called ``thermal'' contributions involving two particles, one
travelling forward in time, the other propagating backward. These
particles can be a $D$ and a $K$ meson, respectively, leading to the
contribution,
\begin{equation}
\langle D|O_i|K\rangle\langle K|O^\dagger_j|D\rangle
e^{-(T-t)m_{K}}e^{-m_{D}t},
\label{eq:twopart}
\end{equation}
which may be significant around $t=T/2$, making the extraction of the
$D_s$ meson and scattering energies less straightforward. If the
overlaps in Eq.~(\ref{eq:twopart}) are of the same order of magnitude
as the leading forward propagating overlaps in Eq.~(\ref{eq:17}) then
at $t=19a$~($17a$) for $T/a=64$~(48) these contributions are of the
order of the statistical errors in the correlator matrix, decreasing
rapidly for smaller $t$.  In the case of two degenerate particles,
Eq.~(\ref{eq:twopart}) reduces to a constant term which can be removed
by taking finite differences, see Ref.~\cite{Helmes:2015gla}.  Here we
choose $t_{\rm max}<19a$~($17a$) for $T/a=64$~(48) to avoid any
significant contribution from thermal states.

\begin{figure*}
\centerline{
\includegraphics[width=.7\textwidth,clip=]{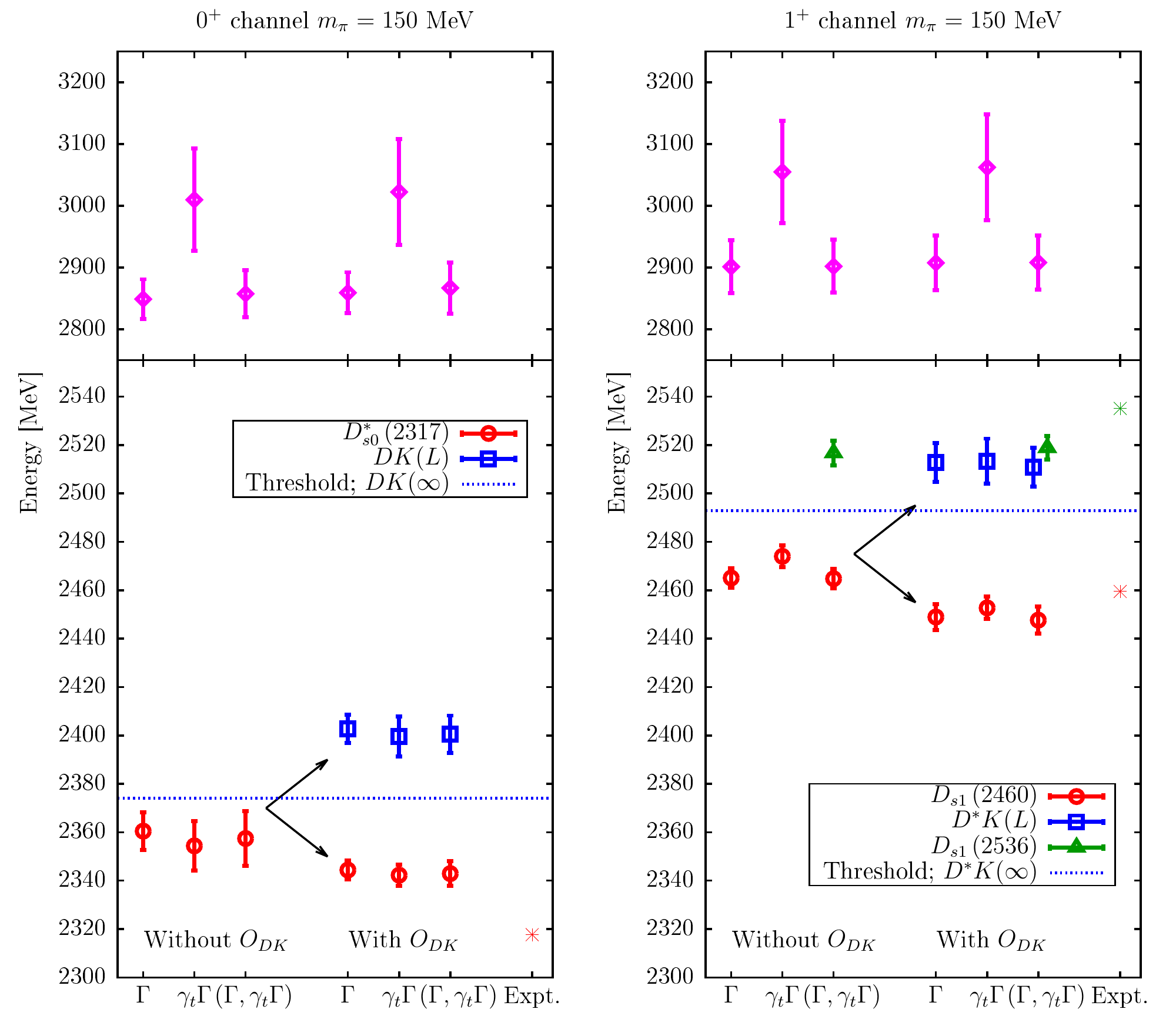}
}
\caption{Lowest energy levels of the $0^+$~(left) and $1^+$~(right)
  channels extracted from fits to the eigenvalues for different
  operator bases for the $m_\pi=150$~MeV, $L/a=64$ ensemble. The
  basis is indicated at the bottom of the figure, where $\Gamma$
  refers to the spin structure of the quark-antiquark interpolators,
  $\bar{s}\Gamma c$~(see Table~\ref{tab_2}) with $\Gamma=\mathbb{1}$ and
  $\gamma_{i}\gamma_{5}$ for the $0^{+}$ and $1^+$ mesons,
  respectively. All smearing levels are utilised for each operator~(see
  Section~\ref{correlator}), such that for the $1^+$ states when
  including the $D^*K$ interpolators, the results labelled with
  ``$\Gamma$'' are determined from a $4\times 4$ correlator matrix,
  while $(\Gamma,\gamma_t \Gamma)$ refers to a $6\times 6$ matrix. The
  exception is the $(\Gamma,\gamma_t \Gamma)$ combination~(with and
  without the $DK$ operators) for the scalar channel for which only
  the $\gamma_t\Gamma$ operator with the largest smearing is employed.
  The non-interacting $DK$ and $D^*K$ thresholds for this ensemble are
  also shown as the dashed blue lines. The black arrows emphasise the
  fact that the lowest energy extracted without the two meson
  operators present is contaminated by contributions from the
  finite volume ``scattering'' state $D^{(*)}K(L)$
  and this level and the ground state are only
  isolated once the $DK$ interpolators are included.}
\label{fig:en_ops}
\end{figure*}

Both single and double exponential fits were performed to each
eigenvalue, giving compatible results as demonstrated in
Fig.~\ref{fig:expvs2exp} for the $m_\pi=290$~MeV, $L/a=64$
ensemble. The starting point for the fit window~($t_{\rm min}$) was set
requiring that the correlated $\chi^2/d.o.f.$ is less than 2 and that
larger values for $t_{\rm min}$ give consistent results within errors.
The energies extracted depend on the operator basis of the correlator
matrix as displayed in Fig.~\ref{fig:en_ops}. In particular, a basis
comprised of only $O_{D_s}$ interpolators gives the first energy level
around 2360~MeV with the next state lying much higher, above
2800~MeV. The $O_{D'_s}$ operators give the same spectrum but with
larger statistical errors for the lowest level, also when 
combined with $O_{D_s}$.

\begin{figure}
\centerline{
\includegraphics[width=.5\textwidth,clip=]{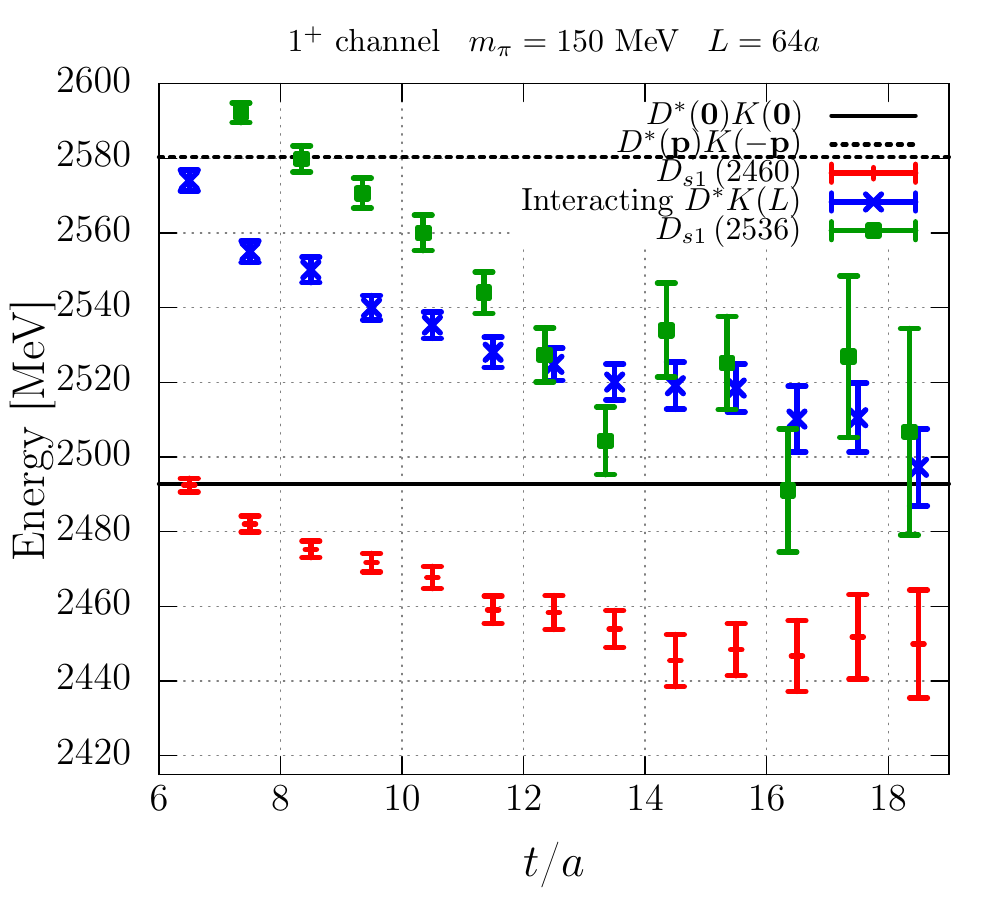}
}
\caption{The effective masses of the first three eigenvalues in the
  axialvector channel for a $6\times 6$ correlator matrix involving a basis
  of $O_{D_s}$, $O_{D'_s}$ and $O_{DK}$ operators for the
  $m_\pi=150$~MeV, $L/a=64$ ensemble. The horizontal lines
  represent the lowest two free scattering states, where for the second level
  corresponding to $D(\boldsymbol{p})K(-\boldsymbol{p})$, the spatial momentum
  $|\boldsymbol{p}|=2\pi/L$.}
\label{fig:effm_onep}
\end{figure}

The first (finite volume) scattering level is only resolved when including the $DK$
operators, with the ground state extracted being shifted approximately
15~MeV lower. This suggests that our choice of two quark interpolators
has overlap with both of the two (closely lying) lowest levels and
that the ground state is not isolated within the time window realised
$t< 19a$ or 1.3~fm if the two meson operators are omitted. We note
that similar observations using two and four quark operator bases
constructed via the distillation approach were made in
Refs.~\cite{Mohler:2013rwa,Lang:2014yfa}, although in general a
different basis, for example, in terms of the spin structure or
spatial extension, can lead to different behaviour. As seen in the
figure, the best signal is obtained from a $4\times 4$ correlator
matrix with all three $O_{D_s}$ operators and the $DK$
interpolator. This turned out to be the case for all ensembles. The
final results for the lowest two levels are summarised in
Table~\ref{tab_4}.

Given the difficulty in extracting the spectrum of closely lying
levels, we remark that the second non-interacting threshold arising
from a $D$ and $K$ meson with opposite momentum,
$|\boldsymbol{p}|=2\pi/L$, lies approximately $85$~MeV above the
first~(with $|\boldsymbol{p}|=0$) for the largest spatial volumes, see
Fig.~\ref{fig:all_effm}. The corresponding finite volume scattering
levels will be similarly close. The inclusion of operators of the form
$D\left(\boldsymbol{p}\right)K\left(-\boldsymbol{p}\right)$~(omitted
in our analysis) would help determine whether the energy of the lowest
scattering level is reliably determined in our analysis.  Any
contamination from higher states is likely to be a small effect,
becoming even less significant for the smaller spatial volumes, as
suggested by the fact that the energy difference between the lowest
two non-interacting thresholds becomes much larger, rising to 494~MeV
for $L/a=24$.

\begin{table}
\caption{Results in MeV for the lowest energy levels extracted in the
  scalar and axialvector channels. The error given is statistical derived
  from jackknife resampling for the chosen fit window. Changing
  the window and/or type of fit~(including one or two exponentials),
  for reasonable $\chi^2/d.o.f$, gives a variation in the central
  values within $\pm 1\sigma$ of the statistical errors. Note that in
  the axialvector case we extract two states (in addition to the scattering level)
  and both are labelled $D_{s1}$. }
\label{tab_4}
\begin{ruledtabular}
\begin{tabular}{cccccc}
 & \multicolumn{2}{c}{$J^P=0^+$} & \multicolumn{3}{c}{$J^P=1^+$}\tabularnewline\hline
$L/a$ & $D_{s0}^*$ & $DK$  & $D_{s1}$  & $D^*K$  & $D_{s1}$\tabularnewline
\hline 
\multicolumn{6}{c}{$m_\pi=290$~MeV}\\\hline
$\:24\:$ & $\:2318(5)\:$ & $\:2594(13)\:$  & $2435(6)$ & $\:2691(16)\:$ & $2549(14)$\tabularnewline
$32$ & $2352(5)$ & $2529(5)$  & $2469(6)$ & $\:2621(14)\:$ & $2540(17)$\tabularnewline
$40$ & $2362(4)$ & $2485(6)$  & $2477(8)$ & $2602(6)$ & $2574(11)$\tabularnewline
$64$ & $2382(3)$ & $2440(5)$ & $2496(4)$ & $2570(3)$ & $2552(5)$\tabularnewline
\hline 
\multicolumn{6}{c}{$m_\pi=150$~MeV}\\\hline
$48$ & $2332(5)$ & $2417(6)$ & $2440(4)$ & $2535(4)$ & $2533(6)$\tabularnewline
$64$ & $2344(4)$ & $2402(6)$ & $2449(5)$ & $2513(8)$ & $2519(5)$
\end{tabular}
\end{ruledtabular}
\end{table}

The analysis of the axialvector channel proceeds in a similar way. In this
case, in addition to the bound state $D_{s1}(2460)$ and scattering level
one expects a resonance, the $D_{s1}(2536)$, just above
threshold. As Figs.~\ref{fig:all_effm} and~\ref{fig:effm_onep} show, an
$O_{D_s}$, $O_{DK}$ basis resolves two closely lying levels, while the
third is only isolated when $O_{D'_s}$ interpolators are included.
Varying the basis for the correlator matrix, we identify the
scattering level to be the one which is only resolved when the $D^*K$
interpolators are included~(like for the scalar channel, see
Fig.~\ref{fig:en_ops} and that tends towards the non-interacting
threshold as the spatial volume increases. The ground state is also
only cleanly extracted when the $D^*K$ interpolators are included, while
for the third level the basis must include both $O_{D_s}$ and
$O_{D'_s}$. The final results for the axialvector channel on all ensembles
are detailed in Table~\ref{tab_4}. In contrast to the ground state,
the third level that we identify as the $D_{s1}(2536)$ is
insensitive to the spatial volume suggesting only a small coupling to the
$D^*K$ threshold. This state lies below the threshold for the ensembles
with $m_\pi=290$~MeV, rising to slightly above but consistent with 
the scattering level for $m_\pi=150$~MeV.

\subsection{Phase shifts, scattering lengths and infinite volume energies}
\label{res:phase}

\begin{figure}
\centerline{
\includegraphics[width=.48\textwidth,clip=]{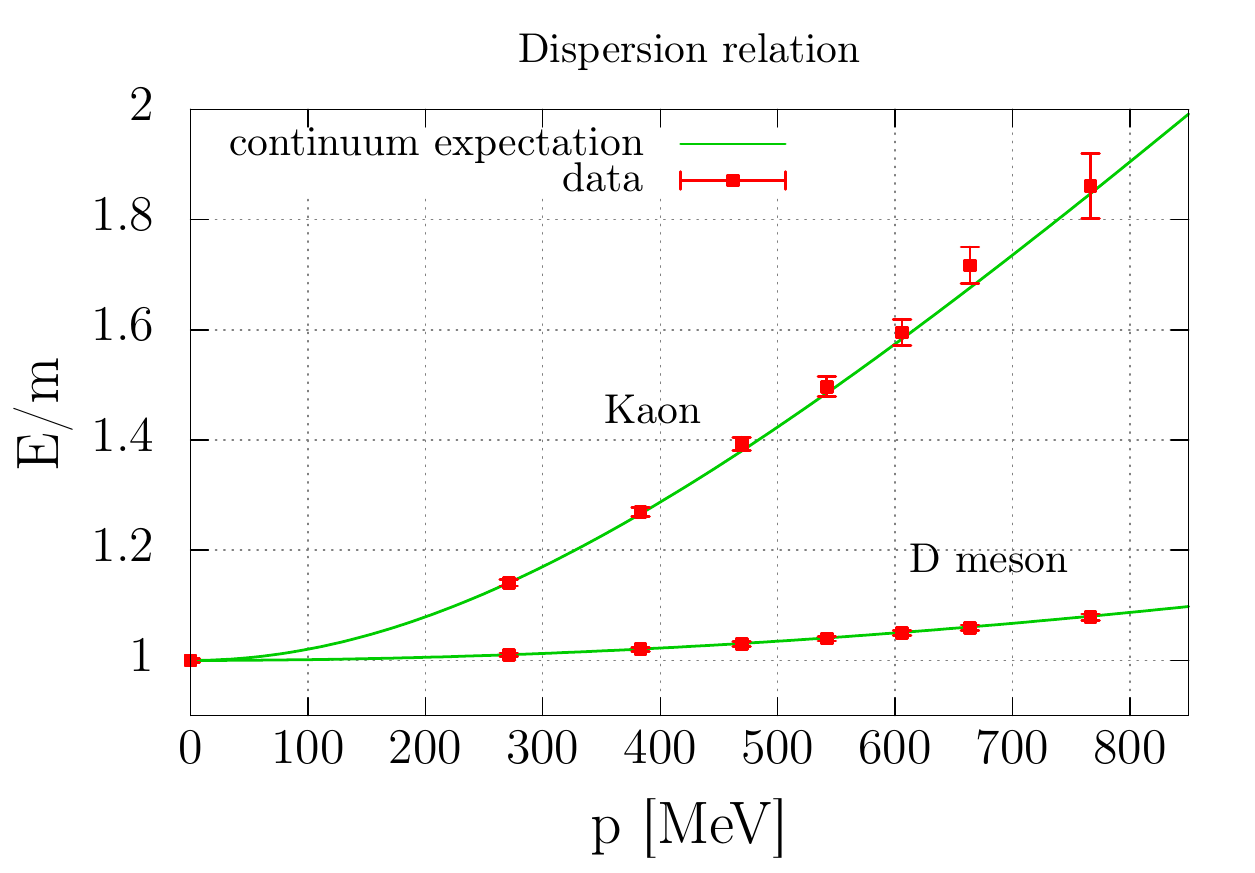}
}
\caption{The dispersion relation for the $K$ and $D$ mesons
  for the $m_{\pi}=150$ MeV, $L/a=64$ ensemble from 
  a subset of configurations, $N_{\rm conf}=600$.}
\label{fig:disp_rel}
\end{figure}

The energy levels presented in the previous subsection are consistent
with the expected spectrum. However, the nature of the physical states
and the infinite volume information --- phase shifts, energies and
scattering lengths etc. --- should be accessed via L\"uscher's relation.
For each energy level we first determine the corresponding momenta of
two particles undergoing elastic scattering via Eq.~(\ref{eq:1}).  The
continuum dispersion relation is assumed to apply for the relevant
$D^{(*)}$ and $K$ mesons, although, discretisation effects can lead to
deviations at finite lattice spacing. Figure~\ref{fig:disp_rel}
demonstrates that the continuum dispersion relation reproduces the
finite momentum $D$ and $K$ meson energies to within the $0.4\%$ and
$0.7\%$ statistical errors, respectively, for the range of momenta of
interest in this study: $p<400$~MeV for the example of 
$m_\pi=150$~MeV and $L/a=64$.  Similar behaviour is seen for
the other ensembles and also for the $D^*$ meson.

The rest masses of the scattering mesons are required as input in
Eq.~(\ref{eq:1}). The values in Table~\ref{tab_1} indicate a mild
dependence on the volume, although this is only statistically
significant~($>3\sigma$) for $m_K$ between $L=24a$ and larger spatial
extents for the $m_\pi=290$~MeV ensembles. We prefer to use the masses
from $L=64a$ as estimates of the infinite volume values throughout
because we are relating the spectra to scattering amplitudes in this
limit. Systematics due to finite $L$ are discussed below.

\begin{figure*}
\centerline{
\includegraphics[width=.8\textwidth,clip=]{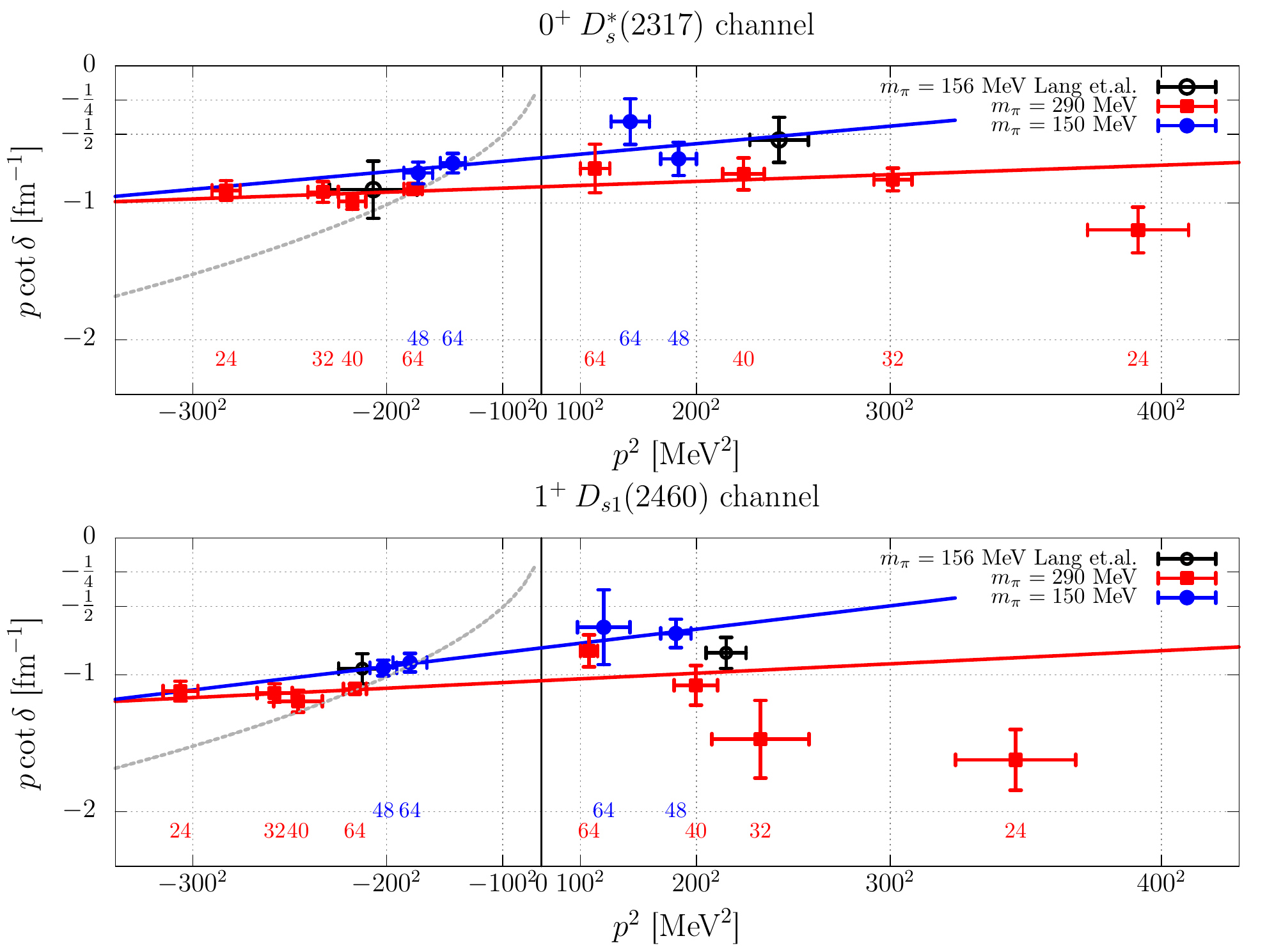}
}
\caption{The combination $p\cot\delta$ as a function of $p^{2}$ for
  the $0^+$ (top) and $1^+$ (bottom) sectors. The threshold $p^{2}=0$ separates the
  bound state (left) and scattering state (right) regions. Linear fits
  to the data excluding the $L=24a$ results are shown as red and blue
  lines while the dashed curve indicates
  $ip=-\sqrt{-p^2}$. The inverse scattering length $1/a_{0}$ can be
  read off from the intersection with the threshold. The results of
  Lang et al.~\cite{Lang:2014yfa} from an ensemble with near physical
  pion mass are shown for comparison.}
\label{fig:pcotd_infvol_withlang}
\end{figure*}

For the ground state and scattering level in the scalar and axialvector
channels, the phase shifts are extracted in the combination
$p\cot\delta$ utilising Eq.~(\ref{eq:2}).  The third state in the
axialvector channel is treated separately due to the lack of volume
dependence, indicating a small coupling to the $D^*K$ threshold.  This
is discussed further in Section~\ref{res:spec}.
Figure~\ref{fig:pcotd_infvol_withlang} presents the results as a
function of $p^{2}$ for all ensembles. The intersection of the data
with the curve representing $ip=-\sqrt{-p^{2}}$ indicates the position
of the pole in the $T$-matrix in infinite volume~(according to
Eqs.~(\ref{eq:12}) and (\ref{eq:14})). As seen in the figure, the
results from the largest ensembles for both channels and pion masses
lie very close to the intersection.

Within the effective range approximation of Eq.~(\ref{eq:3}),
$p\cot\delta$ is linearly dependent on $p^2$. The data are reasonably
consistent with this expectation apart from the results of the
smallest spatial volume, $L=24a\approx 1.7$~fm at $m_{\pi}=290$~MeV.
This may be due to the
breakdown of the approximation and/or the presence of finite volume
effects that are exponentially suppressed with $Lm_\pi$, not taken
into account in L\"{u}scher's formalism.  Performing a linear fit
excluding the $L=24a$ data, we obtain the scattering length $a_{0}$
and the effective range $r_{0}$. The infinite-volume binding momentum,
$p_{B}$, can then be accessed via Eq.~(\ref{eq:16}) and subsequently
the bound state mass and the coupling $g$ through Eqs.~(\ref{eq:1})
and~(\ref{eq:31}), respectively. Note that in terms of $Lm_{\pi}$ the $L=48a$
lattice at $m_{\pi}=150$~MeV is similar in size, however, in this case
$p^2$ is closer to the threshold and to leading order
in ChPT the exponential corrections
are additionally suppressed by a factor of $m_{\pi}^2$.

\begin{table*}
\caption{Scattering length $a_0$, effective range $r_0$,
  infinite-volume binding momentum $|p_B|$, threshold splitting
  $\Delta m$, infinite volume mass $m_{D_s}$ and coupling $g$ for the
  scalar and axialvector channels for the $m_{\pi}=290$ MeV and
  $m_{\pi}=150$ MeV ensembles. The first error is statistical while
  the second indicates the shift in the central value if the analysis
  is repeated using only the $L/a=64$ data for $m_\pi=150$~MeV and
  $L/a=40$ and $64$ data for $m_\pi=290$~MeV. The physical value of
  $m_{D_s}$ and $\Delta m = m_D+m_K-m_{D_s}$ for the QCD theory are
  also given~(labelled as ``Expt''). See Section~\ref{latsetup} for
  details of how isospin breaking and electromagnetic effects are
  taken into account.
}
\label{tab_5}
\begin{ruledtabular}
\begin{tabular}{ccccccc}
                            & \multicolumn{3}{c}{$0^{+}$ channel}    & \multicolumn{3}{c}{$1^{+}$ channel}\tabularnewline
                            &  $m_\pi=290$~MeV & $m_{\pi}=150$~MeV  & Expt. & $m_{\pi}=290$~MeV &  $m_{\pi}=150$~MeV & Expt. \tabularnewline
      \hline 
      $a_{0}$ [fm]          & $-1.13(0.04)(+0.05)$        & $-1.49(0.13)(-0.30)$       &              & $-0.96(0.05)(-0.04)$        & $-1.24(0.09)(-0.12)$         &  \tabularnewline
      $r_{0}$ [fm]          & $0.08(0.03)(+0.08)$       & $0.20(0.09)(+0.31)$       &              & $0.11(0.06)(+0.08)$       & $0.27(0.07)(+0.13)$        &  \tabularnewline
      $|p_{B}|$ [MeV]  & $180(6)(0)$          & $142(11)(-9)$         &              & 
$219(7)(0)$          & $180(11)(-3)$          &  \tabularnewline
      $\Delta m$ [MeV]      & $40(3)(0)$       & $26(4)(-3)$       & $42.6(0.7)(2.0)$       
& $59(4)(0)$       & $42(5)(-2)$        &  $42.9(0.7)(2.0)$  \tabularnewline
      $\:m_{D_{s}}\:$ [MeV] & $\:2384(2)(-1)\:$ & $\:2348(4)(+6)\:$ & $\:2317.7(0.6)(2.0)\:$ 
& $\:2497(4)(-1)\:$ & $\:2451(4)(+1)\:$  &  $\:2459.5(0.6)(2.0)\:$  \tabularnewline
      $g$ [GeV] & $11.9(0.3)(+0.5)$ & $11.0(0.6)(+1.2)$ & $       $ & $14.2(0.6)(+0.7)$ & $13.8(0.7)(+1.1)$  &  $     $ 
      \end{tabular}
\end{ruledtabular}
\end{table*}

\begin{figure*}
\centerline{
\includegraphics[width=.8\textwidth,clip=]{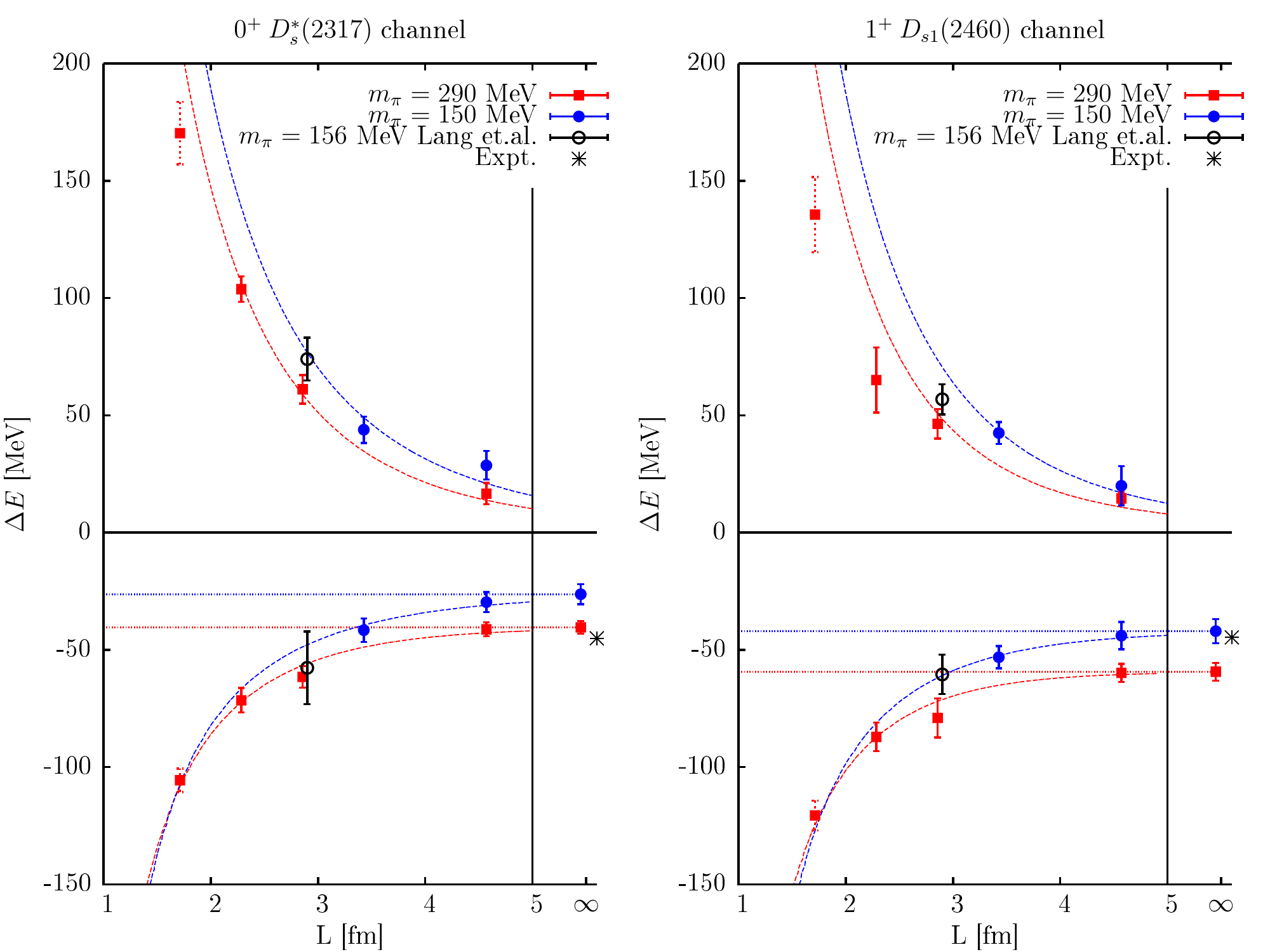}
}
\caption{The splittings of the two lowest states with the
  non-interacting threshold for the scalar and axialvector channels
  for $m_\pi=290$~MeV and 150~MeV. Displayed as dashed lines is the
  dependence on $L$ derived using the effective range approximation
  for $p\cot\delta$ and Eqs.~(\ref{eq:5}) and~(\ref{eq:2}) with the
  central values for $a_0$ and $r_0$ of Table~\ref{tab_4}. The
  infinite-volume splitting, also given in Table~\ref{tab_4}, is
  shown~(statistical errors only) along with the corrected
  experimental values. The horizontal lines indicate the infinite
  volume binding energy of the states for each pion mass. In addition,
  the results of Ref.~\cite{Lang:2014yfa}~(Lang et al.) are included
  for comparison. }
\label{fig:Lp2_infvol}
\end{figure*}

The results for these quantities are compiled in Table~\ref{tab_5}.
The first error given corresponds to the statistical uncertainty while
the second is an estimate of possible residual finite volume effects
due to the exponentially suppressed terms mentioned above. This
estimate is computed by performing the fits to $p\cot\delta$ excluding
the data from the smallest spatial extent. This means using only the
$L/a=64$ results, i.e. two data points, at $m_\pi=150$~MeV and the
$L/a=40$ and $64$ results at $m_\pi=290$~MeV. The shifts in the
central values for most quantities are around one to two statistical
standard deviations or less of the original results. Larger shifts
are found for $a_0$ and $r_0$, in particular, for the lightest
ensemble, however, the results are still consistent given the larger
statistical errors for the reduced fits.

In both channels the scattering length is negative, compatible with
the existence of a bound state. The masses of these states depend on
the pion mass, decreasing by 36(4)~MeV and 46(5)~MeV between
$m_\pi=290$ and 150~MeV for the $0^+$ and $1^+$, respectively. The
errors indicated are due to statistics only.  Similarly, the second
$1^+$ level also decreases by 33(7)~MeV~(see the $L=64a$ data in
Table~\ref{tab_4}).  These shifts are much larger than for the lower
lying pseudoscalar and vector $D_s$ meson masses which decrease by
3~MeV~(from 1980(1)~MeV at $m_\pi=290$~MeV to 1977(1) at
$m_\pi=150$~MeV) and 7~MeV~(from 2101(1)~MeV to 2094(1)~MeV),
respectively, hinting that the $0^+$ and $1^+$ states may have a more
complicated internal structure.  The (lower) axialvector level for the
smallest pion mass is reasonably consistent with experiment, while the
scalar lies somewhat high. This mismatch is likely to be due to
discretisation effects and is discussed further in
Section~\ref{res:spec}. As expected, considering
Fig.~\ref{fig:pcotd_infvol_withlang}, the results for the largest
spatial extent at each pion mass in Table~\ref{tab_4} are consistent
with the infinite volume values.
\begin{figure*}
\centerline{
\includegraphics[width=.7\textwidth,clip=]{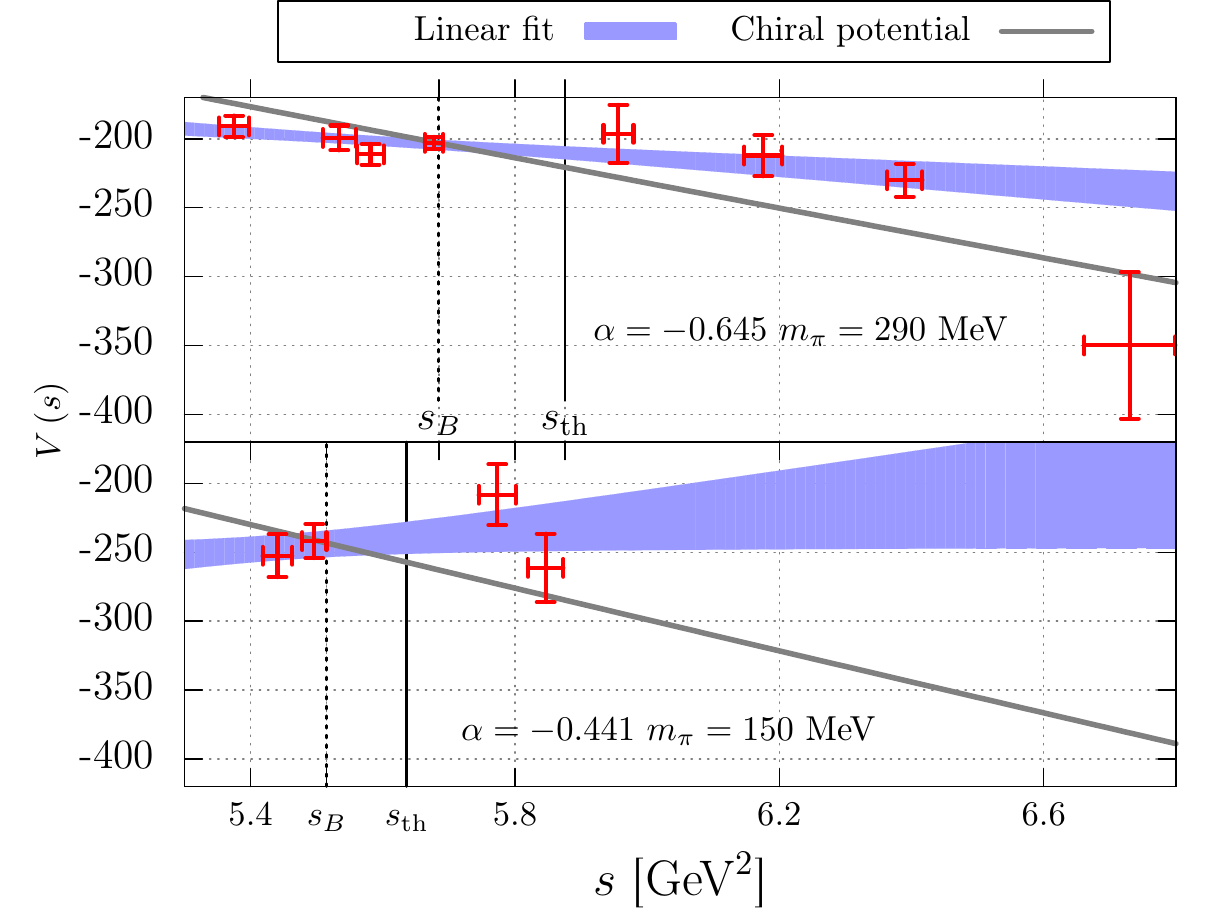}
}
\caption{The potential of the scattering $D$ and $K$ meson in the
  scalar channel as a function of the square of the energy in the
  centre of momentum frame. The subtraction constant $\alpha(\mu)$
  of Eq.~(\ref{eq:45}), utilised for each pion mass, is fixed such that
  the potential from HMChPT~(Eq.~(\ref{eq:38}), also shown as a grey
  line) reproduces the lattice bound state mass for the $L=64a$
  ensembles.  The renormalisation scale $\mu$ is set to $m_D$.  Linear
  fits to the lattice data are shown with one sigma error bars, while
  vertical lines indicate the squared energy of the bound
  state~($s_B$) and also the non-interacting
  threshold~($s_{\rm th}$). Note that the potential is defined to be
  dimensionless. }
\label{fig:pot}
\end{figure*}

A comparison can be made with the study of Ref.~\cite{Lang:2014yfa},
which also includes a near physical pion mass ensemble with
$m_{\pi}=156$ MeV, although the lattice spacing is coarser,
$a=0.09$~fm, and the spatial extent is smaller, $L=2.9$~fm. As shown
in Fig.~\ref{fig:pcotd_infvol_withlang}, the results for $p\cot\delta$
are consistent for both the scalar and axialvector cases, in
particular, when comparing with the linear fit to our data at the
larger $|p^2|$ values realised in Ref.~\cite{Lang:2014yfa}. Not
surprisingly, the scattering lengths and effective ranges they extract
are similar to ours with $a_0=-1.33(20)$~fm and $r_0=0.27(17)$~fm for
the scalar and $a_0=-1.11(11)$~fm and $r_0=0.10(10)$~fm for the
axialvector. The coupling for this simulation was evaluated in a
separate study~\cite{Torres:2014vna} with the results,
$g=12.6(1.5)$~GeV and $12.6(7)$~GeV for the scalar and axialvector
channels, respectively, in reasonable agreement with our values in
Table~\ref{tab_5}. This study focused on an analysis of the Mohler et
al.~\cite{Mohler:2013rwa} and Lang et al.~\cite{Lang:2014yfa} data
within the chiral unitary approach~\cite{Torres:2014vna}, discussed in
the next subsection.

Another quantity of interest is the binding energy, i.e. the splitting
of the bound state with respect to the (non-interacting)
threshold. This is computed at finite $L$ as well as in the infinite
volume limit. The values for the latter~(denoted $\Delta m$) are given
in Table~\ref{tab_4} while the dependence on $L$ is displayed in
Fig.~\ref{fig:Lp2_infvol} together with the results of
Ref.~\cite{Lang:2014yfa} for $m_\pi=156$~MeV for comparison. Also
included in the figure is the same splitting for the lowest scattering
levels, which, as expected, tends to zero with increasing spatial
extent. To guide the eye, we employ the effective range approximation
together with the fits to $p\cot\delta$ shown in
Fig.~\ref{fig:pcotd_infvol_withlang} to derive the dependence on $L$
via Eqs.~(\ref{eq:5}) and~(\ref{eq:2}), indicated by the dashed
lines. The consistency found with the data is a reflection of the
agreement seen in Fig.~\ref{fig:pcotd_infvol_withlang}. For
$m_\pi=150$~MeV, $\Delta m$ in the axialvector channel is compatible
with the physical values, while we undershoot by 17~MeV for the scalar
case.  Taking the spin-average of the two channels to minimise lattice
spacing effects~(see Section~\ref{res:spec}) gives a splitting of
$\Delta \overline{m}=38(4)$~MeV which is within 2$\sigma$ of
$43(7)(2.0)$~MeV for the QCD theory.  We remark that the scalar and
axialvector states are more strongly bound for heavier pion mass.

\subsection{Potential}
\label{res:pot}

We now consider the chiral unitary approach as an alternative method
for extracting the bound state mass and coupling.  The first step is
to compute the potential through Eq.~(\ref{eq:34}) for each energy
level squared $s_n$. We employ dimensional regularisation for the
continuum loop function $G(s)$ for a range of $\alpha(\mu)$ from
$-0.4$ to $-2.2$ with the renormalisation scale fixed to $\mu=m_D$ and
$m_{D^*}$ for the scalar and axialvector cases, respectively. This
range is chosen to encompass values consistent with imposing a cut-off
of $k_{\rm max}\sim\sqrt{\Lambda_\chi^2-m_K^2}\sim 0.87$~GeV in
Eq.~(\ref{eq:gs}), where the chiral symmetry breaking scale
$\Lambda_\chi\sim 1$~GeV. In particular, in Ref.~\cite{Guo:2006fu}
$G(s)$, evaluated by imposing $k_{\rm max}=0.8-0.9$~GeV, was found to be
equivalent to $\alpha\sim -0.6$. The results for the scalar potential
are displayed in Fig.~\ref{fig:pot} for the values of $\alpha$ which
match $V(s)$ for the $L=64a$ ensembles to the HMChPT potential
Eq.~(\ref{eq:38}). In the axialvector case the potential shows a similar
dependence on the squared energy.

\begin{table*}
\caption{The bound state mass, the coupling and the compositeness
  $1-Z$ for the scalar and axialvector channels extracted using the
  chiral unitary approach via linear fits to the potential for a range
  of values of the subtraction constant $\alpha$~(see the text). The
  first error is statistical while the second indicates the shift in
  the central value if the analysis is repeated using only the
  $L/a=64$ data for $m_\pi=150$~MeV and the $L/a=40$ and $64$ data for
  $m_\pi=290$~MeV.}
\label{tab_7}
\begin{ruledtabular}
\begin{tabular}{ccccccc}
&      \multicolumn{6}{c}{Scalar} \\
                            & \multicolumn{3}{c}{$m_{\pi}=290$~MeV}    & \multicolumn{3}{c}{$m_{\pi}=150$~MeV}\tabularnewline
      \hline
                 $\alpha$         & -0.4 & -1.4 & -2.2 & -0.4 & -1.4 & -2.2 \tabularnewline
      \hline 
      $\:m_{D_{s}}\:$~[MeV] & $2384(3)(0)$ & $2384(2)(0)$  & $2384(2)(0)$  
                            & $2348(5)(+3)$ & $2348(4)(+3)$  & $2348(4)(+3)$
\tabularnewline
      $g$~[GeV]  & $11.7(0.3)(+0.7)$ & $11.7(0.3)(+0.7)$ & $11.8(0.3)(+0.7)$ 
                 & $11.2(0.6)(+1.0)$ & $11.1(0.6)(+1.0)$ & $11.1(0.6)(+1.1)$  
 \tabularnewline
      $1-Z$  & $0.90(0.04)(+0.10)$  & $0.90(0.03)(+0.10)$  & $0.90(0.03)(+0.10)$ 
                  & $1.08(0.08)(+0.23)$ & $1.04(0.08)(+0.30)$ & $1.04(0.08)(+0.31)$
 \tabularnewline
      \hline  
&       \multicolumn{6}{c}{Axialvector}\\
& \multicolumn{3}{c}{$m_{\pi}=290$~MeV}    & \multicolumn{3}{c}{$m_{\pi}=150$~MeV}\tabularnewline\hline
$\alpha$ & -0.4 & -1.4 & -2.2 & -0.4 & -1.4 & -2.2 \tabularnewline\hline
$\:m_{D_{s}}\:$~[MeV]      & $2500(4)(-3)$ & $2498(4)(-1)$  & $2497(3)(-1)$  
                            & $2451(4)(+1)$ & $2451(4)(+1)$ & $2451(4)(+1)$\tabularnewline
$g$~[GeV]                 & $14.3(0.5)(+1.2)$ & $14.1(0.5)(+1.0)$ & $14.0(0.5)(+1.0)$ 
                 & $13.8(0.6)(+0.6)$ & $13.8(0.6)(+1.0)$ & $13.8(0.6)(+1.0)$ \tabularnewline
$1-Z$         & $1.00(0.08)(+0.14)$ & $0.95(0.07)(+0.14)$  & $0.94(0.07)(+0.13)$ 
                  & $1.13(0.08)(+0.17)$ & $1.14(0.09)(+0.19)$ & $1.14(0.09)(+0.19)$ 
      \end{tabular}
\end{ruledtabular}
\end{table*}

\begin{table*}
\caption{Comparison of results for the scattering length, effective
  range, coupling and compositeness for the $D_{s0}^*(2317)$ and the
  $D_{s1}(2460)$ from the lattice and unitarised HMChPT. Note that the
  lattice results of this work and Mohler et al.~\cite{Mohler:2013rwa} and Lang et
  al.~\cite{Lang:2014yfa} were obtained using near
  physical pion masses, $m_\pi=150$~MeV and 156~MeV, respectively.
  The $^\dagger$ symbol indicates that the coupling is given in
  Ref.~\cite{Torres:2014vna} where a re-analysis of the data from
  Refs.~\cite{Mohler:2013rwa,Lang:2014yfa} was also performed within the
  effective range approximation. For the
  HMChPT studies we indicate if lattice and/or experimental input has
  been utilised, see the references for details. Liu et al.\ in
  Ref.~\cite{Liu:2012zya} perform a lattice study of $K\overline{D}$
  at unphysical quark mass and use SU(3) flavour symmetry to relate
  the results to that for the $DK$ system.}
\label{tab:comp}
\begin{ruledtabular}
\begin{tabular}{lcccc}
 & $a_0$~[fm] & $r_0$~[fm] & $g$~[GeV] & $1-Z$ \\\hline
\multicolumn{5}{c}{Scalar}\\\hline
This work & -1.49(0.13)(-0.30) & 0.20(0.09)(+0.31) & 11.0(0.6)(+1.2) & 1.04(0.08)(+0.30)\\
Refs.~\cite{Mohler:2013rwa,Lang:2014yfa}:  LQCD & -1.33(20) & 0.27(17) & 12.6(1.5)$^\dagger$ &\\\hline
Ref.~\cite{Torres:2014vna}: HMChPT+LQCD~\cite{Mohler:2013rwa,Lang:2014yfa} & -1.3(5)(1) & -0.1(3)(1) &  11.3 & 0.72(13)(5)\\
Ref.~\cite{Liu:2012zya}: LQCD+HMChPT & -0.86(3) &  & & 0.72-0.66 \\
Ref.~\cite{Guo:2006fu}: HMChPT+Expt &  &  & 10.203 &\\
Ref.~\cite{Yao:2015qia}: HMChPT+Expt+LQCD~\cite{Liu:2012zya,Mohler:2013rwa,Lang:2014yfa} & $-1.04^{+0.06}_{-0.03}$ &  & &\\
Ref.~\cite{Guo:2015dha}: HMChPT+Expt+LQCD~\cite{Liu:2012zya,Mohler:2013rwa,Lang:2014yfa} & $-0.89^{+0.06}_{-0.10}$ & &  &\\
Ref.~\cite{Albaladejo:2016hae}: HMChPT+Expt & $-0.95^{+0.15+0.08}_{-0.15-0.13}$ &  &  & $0.70^{+4+4}_{-6-8}$\\\hline 
\multicolumn{5}{c}{Axialvector}\\\hline
This work & -1.24(0.09)(-0.12) & 0.27(0.07)(+0.13) & 13.8(0.7)(+1.1) & 1.14(0.09)(+0.19)
\\
Refs.~\cite{Mohler:2013rwa,Lang:2014yfa}: LQCD & -1.11(11) & 0.10(10) & 12.6(7)$^\dagger$ &  \\\hline
Ref.~\cite{Torres:2014vna}: HMChPT+LQCD~\cite{Mohler:2013rwa,Lang:2014yfa} & -1.1(5)(2) & -0.2(3)(1) &  14.2 & 0.57(21)(6)
\end{tabular}
\end{ruledtabular}
\end{table*}

The next step is to fit the potential with a reasonable functional
form. A linear ansatz is the natural choice in the small region around
threshold we are considering and is consistent with the data, apart
from the smallest volume ensemble at $m_\pi=290$~MeV. For the latter,
we may be observing finite volume effects, although there is also the
possibility of the influence of the $D_s\eta$ threshold or
Castillejo-Dalitz-Dyson poles~\cite{Castillejo:1955ed}.
Performing linear fits~(omitting the $L=24a$ results) and utilising
Eqs.~(\ref{eq:44}) and (\ref{eq:33}) we obtain the bound state masses
and couplings given in Table~\ref{tab_7}. These physical results are
independent of the subtraction constant employed, as they should be,
and are compatible with the values determined through L\"uscher's
formalism and the effective range approximation.  The two errors shown are,
respectively, statistical and systematic,
representing an estimate of finite volume effects, computed by
performing a reduced fit in the same way as discussed in the previous
subsection.  Note that the phase shift extracted in this approach
through Eqs.~(\ref{eq:42}) and (\ref{eq:12}) is numerically very
similar to the results of the previous subsection and hence the
effective range and scattering length extracted are in agreement with
the values in Table~\ref{tab_5}.

For comparison we also display the scalar potential from leading order
HMChPT~\cite{Kolomeitsev:2003ac} in Fig.~\ref{fig:pot}. We apply the
values of $m_D$, $m_K$ and $F_\pi$ from the $L=64a$ ensemble for each
pion mass. The pion decay constant, determined in
Ref.~\cite{Bali:2014nma}, is equal to 95.1(3)~MeV at $m_\pi=290$~MeV and
85(1)~MeV at $m_\pi=150$~MeV, indicating that we undershoot the
experimental result. This may be due to discretisation effects at the
present lattice spacing~($a=0.071$~fm). The value of $\alpha$ for each
pion mass is chosen such that the bound state energy level for the
largest ensemble is reproduced by the HMChPT potential.  This matching
is reflected in the figure by the potential intersecting the large
ensemble results. One can see that for the short range of $s$ realised
in the lattice data this potential is approximately linear.  The slope
is somewhat steeper than the lattice data suggests and the couplings
derived from Eqs.~(\ref{eq:33}) and~(\ref{eq:38}), $g=10.7$~GeV and
9.8~GeV for $m_\pi=290$~MeV and 150~MeV, respectively~(that are
independent of the subtraction constant) are slightly lower compared
to the results from our fits, cf. Table~\ref{tab_7}. If the
phenomenological values for the masses and decay constant are utilised,
the HMChPT potential gives $g=10.7$~GeV. 

Details of the higher order HMChPT terms for the potential can be
found in
Refs.~\cite{Hofmann:2003je,Guo:2008gp,Guo:2009ct,Cleven:2010aw,Guo:2015dha,Yao:2015qia}
and of other chiral models, for example, in
Ref.~\cite{Gamermann:2006nm}. These works also consider coupled
channel effects. Table~\ref{tab:comp} compares recent results
employing HMChPT with this study and that of Mohler et
al.~\cite{Mohler:2013rwa} and Lang et al.~\cite{Lang:2014yfa}, where
most works determine the scattering length. In many cases some input
from the lattice is taken and overall $a_0$ tends to be lower.

Regarding the compositeness of the bound state, we find a strong $DK$
component in the wave function with $1-Z\approx 1$ to
within 2 sigma in the statistical errors for $m_\pi=150$~MeV for both
the scalar and axialvector channels, with slightly lower values for
the larger pion mass. A large systematic shift is encountered when
trying to estimate finite volume effects, in particular, for
$m_\pi=150$~MeV due to the limited number of data points available.
These results are higher than those determined in a similar analysis
of the Mohler et al.~\cite{Mohler:2013rwa} and Lang et al.~\cite{Lang:2014yfa} data at
$m_\pi=156$~MeV. The authors of Ref.~\cite{Torres:2014vna} found
$1-Z=0.72(13)(5)$ for the $0^+$ and 0.57(21)(6) for the $1^+$,
although the errors are large.

Finally, HMChPT at leading order provides broadly similar values in
the scalar case which increase with pion mass, with $1-Z = 0.75$ and
0.81 for $m_\pi=290$ and 150~MeV, respectively~(independent of
$\alpha(\mu)$). This can be compared to $1-Z=0.71$ when imposing the
physical values of $F_\pi$, $m_K$ and $m_D$. The HMChPT potential has
also been employed to fit the experimental $DK$ invariant mass
distributions of $B\to DDK$ and $B_s\to\pi DK$ decays, giving a
prediction for $1-Z$ of
$0.70^{+4+4}_{-6-8}$~\cite{Albaladejo:2016hae}. As already 
remarked below Eq.~(\ref{eq:33}), the precise meaning of $Z$ in a
relativistic quantum field theory is not clear.

\subsection{Final spectrum}
\label{res:spec}

\begin{table}
\caption{Final results for the masses, thresholds and splittings of
  the lower lying positive and negative parity $D_s$ spectrum, see
  the text for definitions.  The values for the energies of the negative
  parity states, $D^{(*)}$, $K$ and the $1^{\prime +}$
  state~(identified as the $D_{s1}(2536)$) are taken from the
  $m_\pi=150$, $L=64a$ ensemble and the errors indicated are
  statistical only. The masses of the $0^+$ and $1^+$ correspond to
  the infinite volume values for the near physical pion mass detailed
  in Table~\ref{tab_5}. In these cases both statistical and
  systematic~(due to finite volume effects) uncertainties are
  given. The experimental values provided have been corrected for
  isospin and QED effects, see Section~\ref{latsetup} for details. }
\label{tab:res}
\begin{ruledtabular}
\begin{tabular}{ccc}
         & Energy [MeV]  & Expt [MeV] \\\hline
$m_{0-}$ & 1976.9(2)   &     1966.0(4)       \\
$m_{1-}$ & 2094.9(7)  &     2111.3(6)       \\
$m_{0^+}$ & 2348(4)(+6)  &     2317.7(0.6)(2.0)       \\
$m_{1^{+}}$ & 2451(4)(+1)  &     2459.5(0.6)(2.0)       \\
$m_{1^{\prime +}}$ & 2519(5)  &     2535.1(0.1)(2.0)       \\
$m_{D}+m_K$ & 2374(2)  &   2360.3(4)         \\
$m_{D^*}+m_K$ & 2493(3)  &  2502.4(4)          \\
$m_{-}$ & 2065.4(5)  &     2075.0(4)       \\
$m_{+}$ & 2425(4)(+2)  &     2424.1(0.5)(2.0)       \\
$\frac{1}{4}(m_{D}+3m_{D^*})+m_K$ & 2463(2)  & 2466.8(3)            \\
$m_{1^-}-m_{0^-}$ & 118(1)   &  145.3(7)          \\
$m_{1^+}-m_{0^+}$ & 103(6)($^{+1}_{-6}$)   &  141.8(0.9)(2.0)          \\
$m_{0^+}-m_{0^-}$ & 371(4)(+6)   & 351.7(0.7)(2.0)       \\
$m_{1^+}-m_{1^-}$ & 356(4)(+1)   & 348.2(0.8)(2.0)          \\
$m_{1^{\prime +}}-m_{1^-}$ & 424(5) &  423.8(0.6)(2.0)          \\
$m_{+}-m_{-}$ &  360(3)(+2)   &  349.1(0.6)(2.0)        
\end{tabular}
\end{ruledtabular}
\end{table}

\begin{figure}
\centerline{
\includegraphics[width=.5\textwidth,clip=]{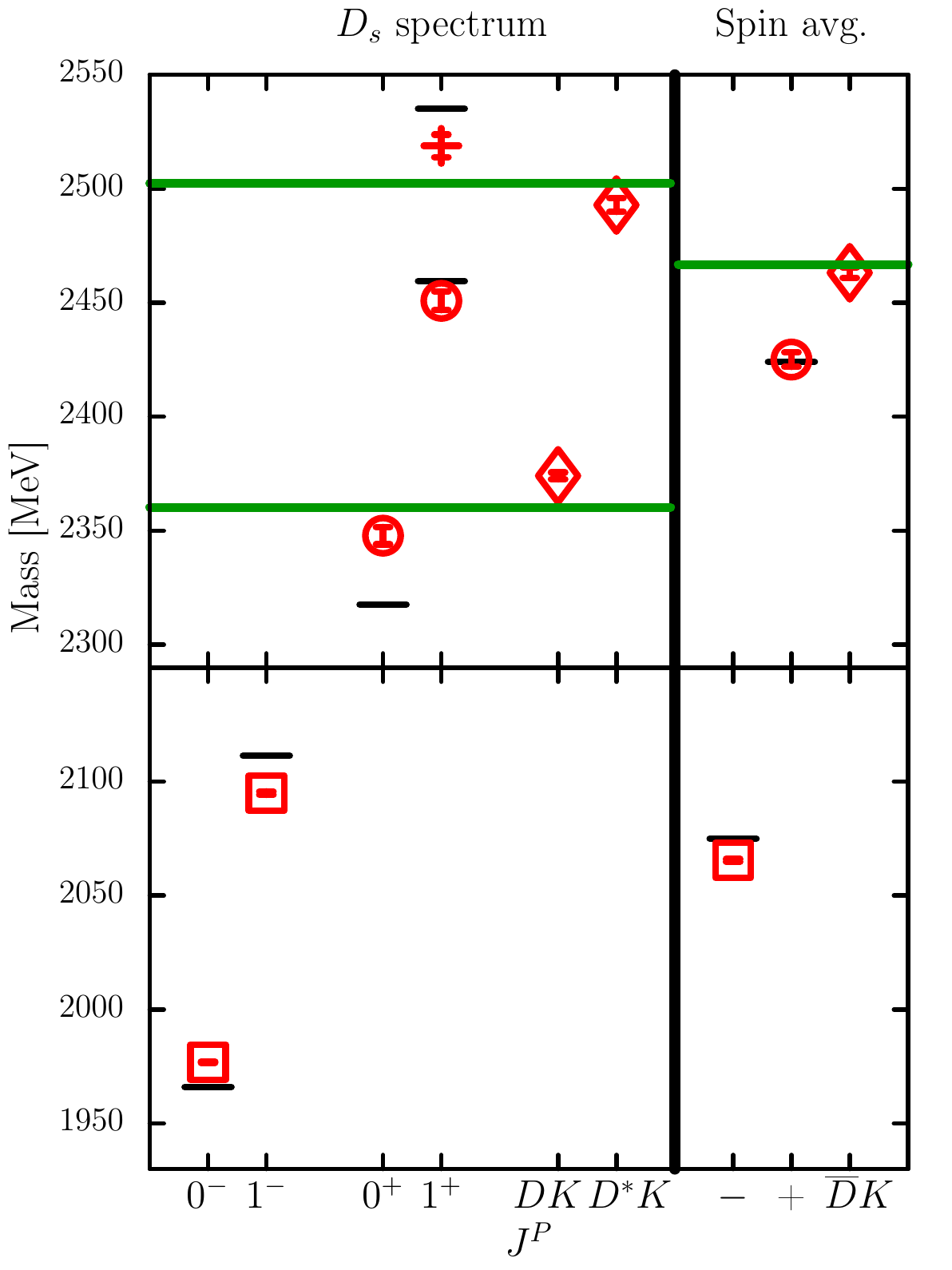}
}
\caption{On the left, our final results for the lower lying $D_{s}$
  spectrum as detailed in Table~\ref{tab:res}. The short horizontal
  black lines indicate the corrected experimental values~(see
  Section~\ref{latsetup}) while the green horizontal lines give the
  positions of the $DK$ and $D^*K$ non-interacting thresholds. Our
  lattice results for the finite volume thresholds are labelled
  $DK$ and $D^*K$, respectively. The errors indicated are
  statistical only. On the right, the negative parity spin-averaged
  $1S$ mass $m_-=\frac{1}{4}\left(m_{0^-}+3m_{1^-}\right)$ is shown
  and denoted $-$, while the same spin-average of the positive parity
  $0^+$ and $1^+$ states is labelled with $+$ and the weighted average
  of the threshold is labelled as $\overline{D}K$.}
\label{fig:allDs_spectra}
\end{figure}

Our final results for the lower lying $D_s$ spectrum are compiled in
Table~\ref{tab:res} and displayed in Fig.~\ref{fig:allDs_spectra}.
The energies of the negative parity particles and the thresholds,
which display very little dependence on the spatial volume, are taken
from the $m_\pi=150$~MeV, $L=64a$ ensemble. The masses of the
$D_{s0}^*(2317)$ and $D_{s1}(2460)$ correspond to the infinite volume
values in Table~\ref{tab_5} derived from the phase shift analysis of
Section~\ref{res:phase}. For the $1^+$ state above threshold,
identified as the $D_{s1}(2536)$, we also found no significant
dependence of the mass on the spatial extent, even in the presence of
$s$-wave $D^*K$ interpolators. This behaviour suggests a small
coupling to the threshold~(which is difficult to resolve on the
lattice via L\"uscher's formalism) and a narrow width. Indeed the
experimentally measured width is only approximately $0.8$~MeV for this
decay mode~\cite{Olive:2016xmw}. It would be interesting to also
consider coupling to the $D^*K$ in $d$-wave since in the heavy quark
limit this mode is dominant for the $j^P=\frac{3}{2}^+$ doublet of
which the $D_{s1}(2536)$ is part, with the $s$-wave channel
absent~\cite{Isgur:1991wq}~(the opposite holds for the
$j^P=\frac{1}{2}^+$ doublet which contains the $D_{s1}(2460)$). 
Experimentally, the $s$-wave mode dominates and its contribution to
the total width is 0.72(5)(1)~\cite{Balagura:2007dya}.  At present,
our best estimate of the physical $D_{s1}(2536)$ energy is again
provided by the $m_\pi=150$~MeV, $L=64a$ ensemble.

We achieve statistical errors below $0.2\%$ for the positive parity
states and even smaller ones for the negative parity states, due to
the large number of configurations analysed. Although the overall
pattern of energy levels is as expected, at this level of precision,
there are clear discrepancies with the experimental spectrum due to
the remaining systematics arising from lattice spacing effects and the
still unphysical light quark mass. As mentioned in
Section~\ref{latsetup}, fine structure splittings are expected to be
sensitive to discretisation effects~(which begin at $O(a^2)$ in our
study), due to being dominated by high energy scales.  We find the
hyperfine splittings, $m_{D_{s}^{*}}-m_{D_{s}}=118(1)$ MeV and
$m_{1^+}-m_{0^+}=103(6)$~MeV, are well below the QED and isospin
corrected experimental values of $145.3(7)$ and $142(2)$~MeV,
respectively. Spin-averaged combinations are less affected, and better
agreement is seen as illustrated on the right hand side of
Fig.~\ref{fig:allDs_spectra} --- both the positive parity and
threshold averages are reproduced within errors --- indicating most of
the disagreement observed for the individual masses is likely due to
discretisation effects. For the positive parity spin-average we are
computing $m_+=\frac{1}{4}\left(m_{0^+}+3m_{1^+}\right)$ for the
$\frac{1}{2}^+$ doublet, which includes the lower axialvector state.
For the threshold we take the spin-average of the $1S$ $D$
mesons masses, $m_{\overline{D}}=\frac{1}{4}(m_D+3m_{D^*})$, together with the kaon mass.

\begin{figure}
\centerline{
\includegraphics[width=.5\textwidth,clip=]{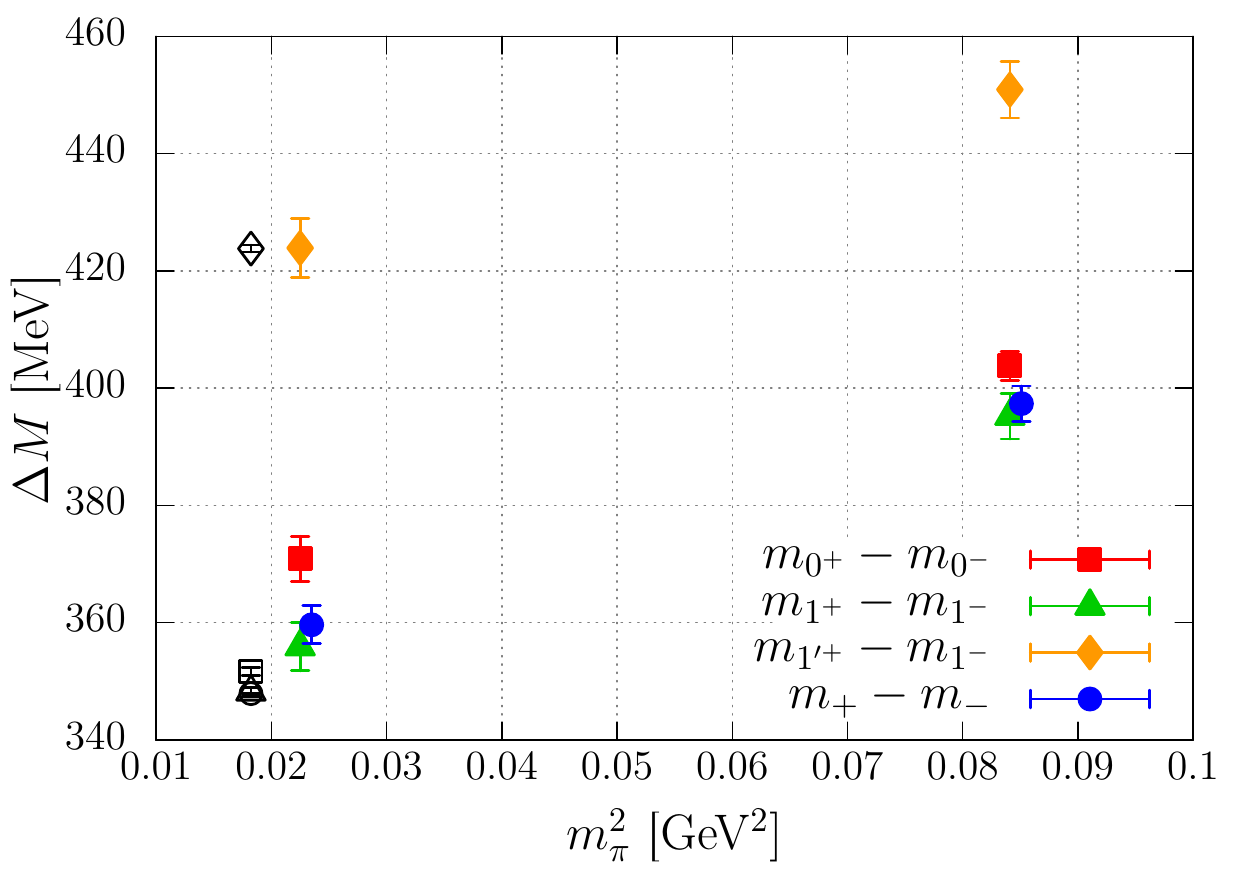}
}
\caption{Mass splittings as a function of the pion mass squared for
  ensembles with a spatial extent of $L=64a$. The corresponding
  corrected experimental values~(see Section~\ref{latsetup}) are
  indicated as black open symbols at $m_\pi=0.135$~GeV. The
  spin-average of the mass of the lowest lying negative~(positive)
  parity states is denoted $m_-$~($m_+$), while
  $m_{1^+}$~($m_{1^{\prime +}}$) denotes the mass of the
  lower~(higher) $1^+$ level. The errors shown are statistical only.}
\label{fig:split}
\end{figure}

In order to separate the light and strange quark effects from that of
the charm quark, we compute the splitting $m_+-m_-$, displayed in
Fig.~\ref{fig:split} for the largest spatial extent. The results for
$m_\pi=290$~MeV are shown for comparison. Heavy quark effects may also
largely cancel when considering splittings between masses within the
two $j=\frac{1}{2}$ doublets, i.e.  $\Delta m_0=m_{0^+}-m_{0^-}$ and
$\Delta m_1=m_{1^+}-m_{1^-}$ and possibly between the lower $j_z$
components of the $\frac{3}{2}^+$ and $\frac{1}{2}^-$ doublets,
$\Delta m_{1^\prime}=m_{1^{\prime +}}-m_{1^-}$. The splittings are a
few hundred MeV in size as expected for quantities dominated by scales
of the order of $\overline{\Lambda}\sim 500$~GeV~($\ll
a^{-1}=2.76$~GeV). As mentioned in Section~\ref{res:phase}, there is
significant dependence on the pion mass which is at odds with a simple
charm-strange quark model interpretation of the positive parity
states~(the masses of the $1S$ negative parity states do not vary
significantly with $m_\pi$). For
$m_\pi=150$~MeV, $\Delta m_1$ and $\Delta m_{1^\prime}$ are reasonably
consistent with experiment, while $\Delta m_{0}$ displays a
significant difference of around $6\%$. However, for the
spin-averaged splitting, for which lattice spacing effects are most
effectively suppressed, there is only a 3\% discrepancy or 4$\sigma$
in the statistical errors. With a very short (crude) linear
extrapolation to the physical point of $m_\pi=135$~MeV, we find
$356(3)$~MeV for this splitting compared to the physical value of
$349(2)$~MeV.

\subsection{Decay constants}
\label{res:decay}

We are interested in how the magnitude of the ground state $0^+$ and $1^+$ decay
constants compare with those of ``conventional'' mesons such as the
pseudoscalar $D_s$ and vector $D^{*}_s$. Starting with the $0^+$
state, the scalar decay constant, $f_S$, is defined through,
\begin{eqnarray}
\braket{0|\overline{s}c|D_{s0}^{*}\left(\boldsymbol{p}\right)} & = & f_{S}m_{0^+}, \label{eq:25}
\end{eqnarray}
where the physical state is normalised according to
\begin{equation}
\braket{D_{s0}^{*}\left(\boldsymbol{p}\right)|D_{s0}^{*}\left(\boldsymbol{p}'\right)}=2E(\boldsymbol{p})L^{3}\delta_{\boldsymbol{p}\boldsymbol{p}'},
\end{equation}
for a finite volume $L^3$ and $E(\boldsymbol{p})$ is the energy of the
state.  The conserved vector current relation~(CVC) connects $f_S$
with the vector decay constant, $f_V$,
\begin{eqnarray}
\braket{0|\overline{s}\gamma_{\mu}c|D_{s0}^{*}\left(\boldsymbol{p}\right)} & = & f_{V}p_{\mu} \label{eq:26},
\end{eqnarray}
such that at zero momentum, 
\begin{equation}
f_V=f_S (m_c-m_s)/m_{D_{s0}^*},\label{eq:29}
\end{equation}
with $m_c$ and $m_s$ denoting the charm and strange quark masses,
respectively.  For a $1^+$ state with polarisation $\epsilon_\mu$, one
can define axialvector and tensor decay constants:
\begin{eqnarray}
\braket{0|\overline{s}\gamma_{\nu}\gamma_5c|D_{s1}\left(\boldsymbol{p},\boldsymbol{\epsilon}\right)} & = & f_{A} m_{D_{s1}} \epsilon_\nu, \label{eq:ax}\\
\braket{0|\overline{s}\gamma_5\sigma_{\mu\nu} c|D_{s1}\left(\boldsymbol{p},\boldsymbol{\epsilon}\right)} & = & f_{T} (p_\mu  \epsilon_\nu -p_\nu  \epsilon_\mu )\label{eq:ten},
\end{eqnarray}
where since we are at zero spatial momentum, we set $\mu=t$ and
average over $\nu=i\in\{1,2,3\}$.  The above normalisations are
compatible with those for a pseudoscalar meson for which the decay
constant $f_{D_s}=250(7)$~MeV for $N_f=2$, see the FLAG
review~\cite{Aoki:2016frl} for details. Note that when comparing with
the latter, the $0^+$ vector and $1^+$ axialvector decay constants are
the corresponding weak observables, while $f_S$ and $f_T$ only appear
in Standard Model processes beyond tree-level or new physics
interactions.

\begin{figure*}
\centerline{
\includegraphics[width=.5\textwidth,clip=]{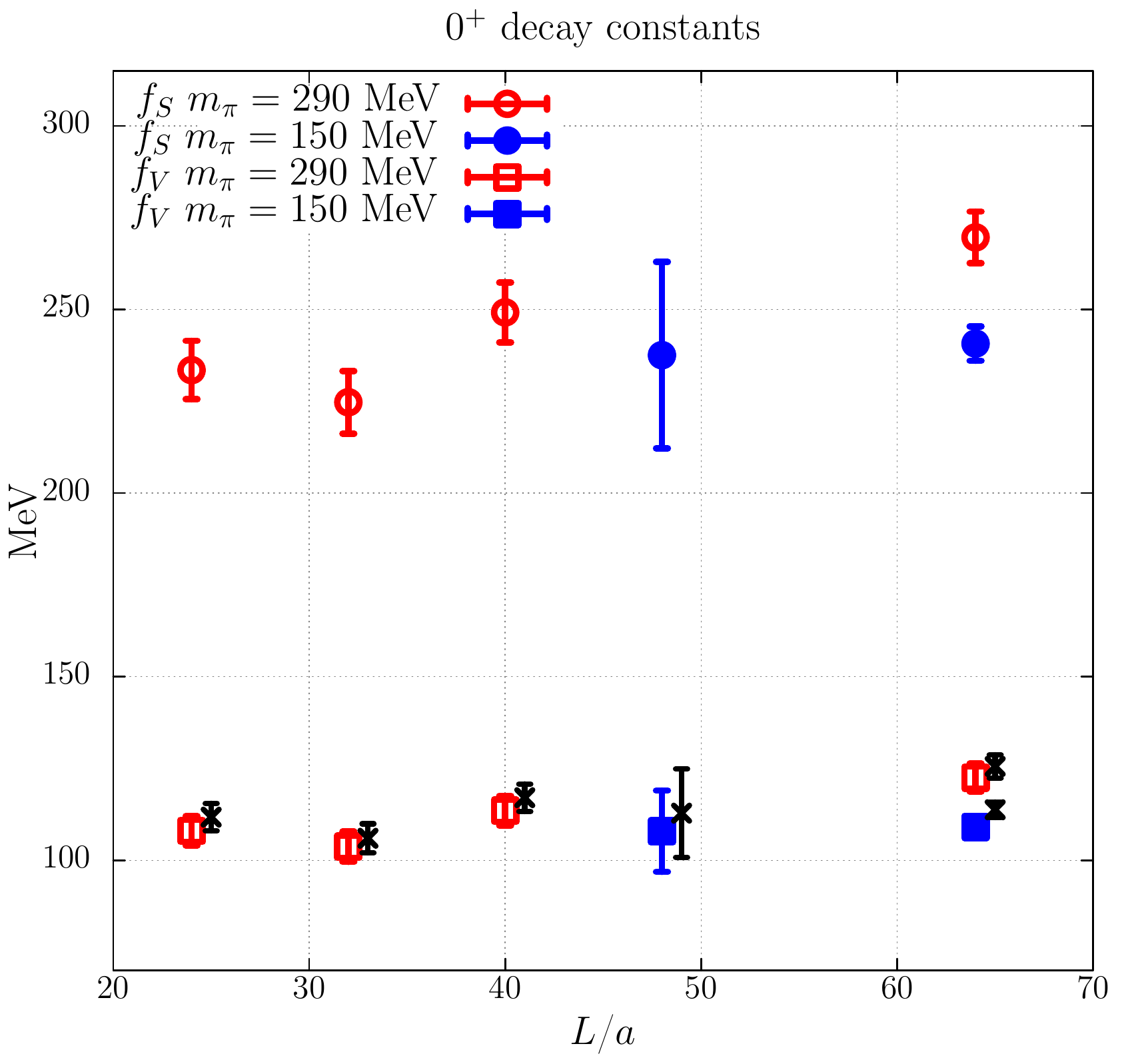}
\includegraphics[width=.5\textwidth,clip=]{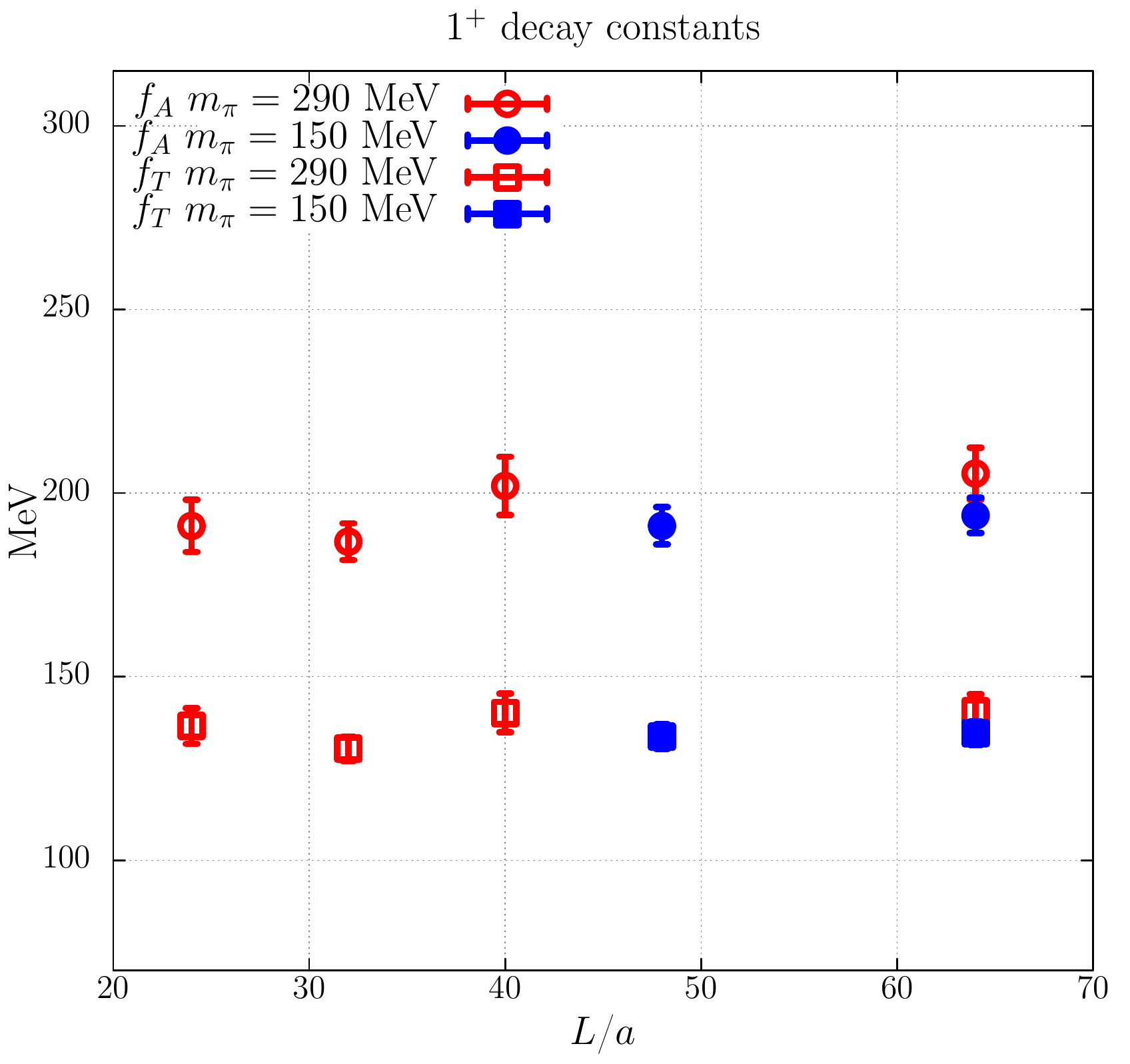}
}
\caption{The scalar and vector decay constants of the
  $D_{s0}^*(2317)$~(left) and the axial and tensor decay
  constants of the $D_{s1}(2460)$~(right) for different pion masses
  and spatial volumes. The black crosses indicate $f^{\rm CVC}_V$, the
  vector decay constant obtained using $f_S$ and the CVC relation,
  Eq.~\eqref{eq:29}. The errors shown correspond to the statistical
  and renormalisation uncertainties added in quadrature, see
  Table~\ref{tab_6}.}
\label{fig:dc_renorm}
\end{figure*}

On the lattice, the bare matrix elements are extracted from
correlators with a source interpolator, $O^\dagger$, which has a
good overlap with the physical state, and local sink operators,
$J_S=\overline{s}c$ and $J_{V}=\overline{s}\gamma_{t}c$ for the $0^+$
and $J_A=\overline{s}\gamma_i\gamma_5 c$ and
$J_{T}=\overline{s}\gamma_5\gamma_{t}\gamma_i c$ for the $1^+$, that are
projected onto zero momentum:
\begin{eqnarray}
C^{X}_{LS}(t) & =
&\braket{0|J_{X}\left(t\right)O^{\dagger}(0)|0}\\ &\approx
&\sqrt{\frac{m L^{3}}{2}e^{m t_{0}}}f^{\rm latt}_{X}e^{-m t} \label{eq:27}
\end{eqnarray}
with $X\in\{S,V,A,T\}$ and $m\in\{m_{0^+},m_{1^+}\}$.  The source
interpolator is constructed from the basis of smeared operators
realised for the variational analysis, weighted by the components of
the eigenvector of the lowest state.  In the limit of ground state
dominance, we expect the time dependence shown on the r.h.s., where
$t_0$ is the reference time in Eq.~(\ref{eq:18}).  We perform
simultaneous single exponential fits to correlators containing
operators with the same quantum numbers, i.e. $C^S_{LS}$ and
$C^V_{LS}$ for the $0^+$ and $C^A_{LS}$ and $C^T_{LS}$ for the $1^+$
$D_s$ mesons.  This ensures the mass in Eq.~(\ref{eq:27}) is
consistent for the different decay constants. The resulting masses
were also found to be compatible with those extracted from the
variational analysis. The correlators relevant for determining the
axial and tensor decay constants of the $D_{s1}(2536)$ were also
computed in our analysis, however, the simultaneous fits were
unsatisfactory and it was not possible to achieve reliable
results. For this reason, we do not present values for the decay
constants of this resonance.

\begin{table*}
\caption{Renormalised decay constants for the
  $D_{s0}^{*}\left(2317\right)$ and $D_{s1}(2460)$ in MeV for all
  ensembles. The scalar and tensor decay constants are renormalised in
  the $\overline{\rm MS}$ scheme at a scale of 2~GeV. The errors given
  are in the first case statistical and in the second case due to the
  uncertainty in the renormalisation and improvement factors. For the $m_\pi=150$~MeV
  data the third error is an estimate of finite volume effects while the
  fourth is the possible order of magnitude of the discretisations
  effects, see the text.}
\label{tab_6}
\begin{ruledtabular}
\begin{tabular}{ccccccl} 
 & \multicolumn{4}{c}{$m_\pi=290$~MeV} & \multicolumn{2}{c}{$m_\pi=150$~MeV}\\
$L/a$ & 24 & 32 & 40 & 64 & 48 & 64\\\hline
\multicolumn{7}{c}{$D_{s0}^*$}\\\hline
$f_{S}^{\rm ren}$ [MeV] & $\:233(8)(2)\:$ &$225(8)(2)$ &$249(8)(2)$ & $270(7)(2)$ & $238(25)(2)$ & $241(4)(2)(+12)(10)$ \tabularnewline
$f_{V}^{\rm ren}$ [MeV]  &  $\:108(3)(2)\:$&  $104(4)(2)$ &  $114(3)(2)$ &  $123(3)(2)$ &  $109(11)(2)$&  $111(2)(2)(+05)(10)$ \tabularnewline
$f^{\rm CVC, ren}_{V}$ [MeV]  &  $\:112(4)(0)\:$&  $106(4)(0)$ &  $117(4)(0)$ &  $126(3)(0)$ &  $113(12)(0)$&  $114(2)(0)(+05)(10)$\tabularnewline\hline
\multicolumn{7}{c}{$D_{s1}$}\\\hline
$f_{A}^{\rm ren}$ [MeV]  & $\:191(6)(4)\:$ &$187(3)(4)$ &$202(7)(4)$ & $205(6)(4)$ & $191(4)(4)$ & $194(3)(4)(+5)(10)$ \tabularnewline
$f_{T}^{\rm ren}$ [MeV]  &  $\:137(4)(2)\:$&  $130(2)(2)$ &  $140(5)(2)$ &  $141(4)(2)$ &  $134(2)(2)$&  $135(2)(2)(+3)(10)$ 
\end{tabular}
\end{ruledtabular}
\end{table*}

In order to convert the bare results, $f_X^{\rm latt}$, into
physical predictions the lattice decay constants are renormalised in
the $\overline{\rm MS}$ scheme and Symanzik improvement is applied to
reduce the discretisation errors to $O(a^2)$,\footnote{In addition to
  employing a non-perturbatively $O(a)$ improved fermion action.}
\begin{equation}
f_X^{\rm ren}=Z_X\left(1+a\overline{m}b_X\right)f^{\rm latt}_X, \label{eq:28}
\end{equation}
where $\overline{m}=(m_c+m_s)/2$ and the vector Ward identity quark
masses, $m_{q=c,s}=(1/\kappa_q-1/\kappa_{\rm crit})/2a$. The critical
hopping parameter, $\kappa_{\rm crit}=0.1364281(12)$, was evaluated in
Ref.~\cite{Bali:2014nma}, which also provides non-perturbative values
for the renormalisation factors, 
\begin{eqnarray}
&&Z_A=0.76487(64), \qquad Z_V=0.7365(48), \nonumber\\
&&Z_S=Z^{\overline{\text
    MS}}_S(\mu=2\,\text{GeV})=0.6153(25),\nonumber\\
&& Z_T=Z^{\overline{\text
    MS}}_T(\mu=2\,\text{GeV})=0.8530(25),
\end{eqnarray} that are updates of earlier
determinations in Ref.~\cite{Gockeler:2010yr}. One loop expressions
for the improvement factors $b_{A,V,T}$ were
employed~\cite{Sint:1997jx,Taniguchi:1998pf,Capitani:2000xi},
\begin{eqnarray}
&& b_A=1+0.15219(5)g^2,\qquad b_V=1+0.15323(5)g^2,\nonumber\\
&& b_T=1+0.1392(1)g^2,
\end{eqnarray}
along with the ``improved'' coupling $g^2=-3\ln P = 6/\beta+O(g^4)$.
$P$ denotes the plaquette with the normalisation $P=1$ at
$\beta=\infty$ and the chirally extrapolated value of $P$ is equal to
$0.54988$. The uncertainty due to omitting higher orders of the
perturbative expansion is taken to be one half of the one-loop term.  For
the scalar case, we utilise the non-perturbative determination of $b_S$
in Ref.~\cite{Fritzsch:2010aw}.

The final results are detailed in Fig.~\ref{fig:dc_renorm} and
Table~\ref{tab_6}.  In the latter, the first error quoted is
statistical, while the second is the uncertainty due to
renormalisation and $O(a)$ improvement. The decay constants tend
to decrease slightly as the pion mass is reduced and for the
$D_{s1}^*(2317)$ there is a mild dependence on the spatial lattice 
extent. We find reasonable consistency with Eq.~(\ref{eq:29}) when we
derive the vector $0^+$ decay constant from the scalar one, as seen in
the figure, suggesting discretisation effects are not severe. We
remark that since the combination $f_S(m_c-m_s)$ is renormalisation
group invariant and is free of additive renormalisation, $f_V$
determined in this way~(denoted $f_V^{\rm CVC}$) does not require
knowledge of any renormalisation factors or improvement terms and is
automatically $O(a)$ improved. We consider $f_V^{\rm CVC}$ to
represent the most reliable estimate of the vector decay constant.

\begin{table*}
\caption{Comparison of lattice results for the scalar and vector decay
  constants of $D_{s1}^*(2317)$ and the axial decay constant of the
  $D_{s1}(2460)$ from this work and that of
  Ref.~\cite{Herdoiza:2006qv} with other approaches, in MeV. The
  errors indicated for our values are, in order, statistical, those
  arising from the renormalisation and estimates of the uncertainties
  due to finite volume and lattice
  spacing. Refs.~\cite{Hwang:2004kga,Cheng:2003kg,Cheng:2006dm}
  combine the experimental branching fractions for $B\to
  D^{(*)}D_{sJ}^{(*)}$ decays with heavy quark symmetry~(HQS) and the
  factorisation approximation, while
  Refs.~\cite{LeYaouanc:2001ma,Hsieh:2003xj,Cheng:2003sm,Verma:2011yw,Segovia:2012yh}
  employ quark models~(QM) and
  Refs.~\cite{Colangelo:2005hv,Wang:2015mxa} use QCD sum
  rules~(QCDSR). The study of Ref.~\cite{Faessler:2007cu} assumes a
  $D^{(*)}K$ molecular structure for the $D_{s0}^*(2317)$ and
  $D_{s1}(2460)$ and constrains the parameters of their effective
  Lagrangian with the experimental $D\rightarrow K^{(*)}$ semileptonic
  formfactors.  See the references for more details.}
\label{tab:comp2}
\begin{ruledtabular}
\begin{tabular}{lcccc}
          & $f_{S}^{0^+}$~[MeV] & $f_{V}^{0^+}$~[MeV] & $f_{A}^{1^+}$~[MeV] \\\hline
This work &   241(4)(2)(+12)(10)            &   114(2)(0)(+5)(10)            &     194(3)(4)(+5)(10)             \\
LQCD~\cite{Herdoiza:2006qv}
          &  340(110)     &  200(50)      &                             \\\hline
$B$-decays+HQS~\cite{Hwang:2004kga}
          &               &  74(11)       &   166(20)                   \\
$B$-decays+HQS~\cite{Cheng:2003kg} 
          &               &  67(13)       &                             \\
$B$-decays+HQS~\cite{Cheng:2006dm}
          &               & 58-86         &   130-200                   \\\hline
QM~\cite{LeYaouanc:2001ma} 
          &               &  440          &    410                      \\
QM~\cite{Hsieh:2003xj} 
          &               &  122-154      &                             \\
Light Front QM~\cite{Cheng:2003sm} 
          &               & 71            &  117                        \\
Light Cone QCDSR~\cite{Colangelo:2005hv} 
          &    225(25)    &               &  225(25)                    \\
$DK$-molecule~\cite{Faessler:2007cu} 
          &               &  67.1(4.5)    &  144.5(11.1)                \\
Light Front QM~\cite{Verma:2011yw} 
          &               &  $74.4^{+10.4}_{-10.6}$ 
                                          &   $159^{-36}_{+32}$       \\
QM~\cite{Segovia:2012yh}
          &               &  119          &  165                   \\
QCDSR~\cite{Wang:2015mxa} 
          &  333(20)      &               &   245(17)                   
\end{tabular}
\end{ruledtabular}
\end{table*}

We take the results from the $m_\pi=150$~MeV, $L=64a$ ensemble as
being closest to the physical values. Unfortunately, the correlators
needed to evaluate the negative parity equivalents were not computed,
however, a simulation with the same action by the ALPHA collaboration
found the pseudoscalar decay constant $f_{D_s}\sim
257$~MeV~\cite{Heitger:2013oaa} at $m_\pi=190$~MeV and $a=0.065$~fm
with a final continuum, chirally extrapolated value of
$247(5)(5)$~MeV. Very little dependence on the pion mass was
observed. Considering this result and the FLAG value quoted above, the
($P$-wave) $0^+$ vector decay constant is roughly $45\%$ of that of
the pseudoscalar, slightly above the estimate of $\sim 0.32$ from
non-leptonic $B$ decays to $D^{(*)}D_{sJ}^{(*)}$ but of a similar
order of magnitude. The difference is indicative of the size of
$1/m_c$ corrections and/or violations of the factorisation
approximation in the latter approach.

Performing the same comparison for the $D_s^*$ and $D_{s1}(2460)$ is
more difficult as lattice results for the vector meson are only
available after continuum and chiral extrapolation for different
lattice actions: Becirevic et al.\ utilising $N_f=2$ twisted mass
fermions found $f_{Ds^*}=311(9)$~MeV and
$f_{Ds^*}/f_{D_s}=1.26(3)$~\cite{Becirevic:2012ti}, while for $N_f=2+1+1$ 
HPQCD with the HISQ fermion action obtained
$f_{Ds^*}/f_{D_s}=1.10(2)$~\cite{Donald:2013sra} and the ETM
collaboration with twisted mass fermions quoted
$f_{Ds^*}=268.8(6.6)$~MeV and
$f_{Ds^*}/f_{D_s}=1.087(20)$\cite{Lubicz:2016bbi}.  Taking
$f_{Ds^*}/f_{D_s}$ in the range $1.1-1.3$ and our result for
$f_{D_{s1}(2460)}$, gives the latter very roughly as $60-70\%$ of
$f_{D_s^*}$, which is very similar to the estimate from non-leptonic
$B$ decays.

With a statistical precision of less than $2\%$ one might expect the
systematics arising from finite volume and discretisation effects to
be noticeable. We quantify the former by performing a finite volume
extrapolation of the $m_\pi=290$~MeV data, where we have a sufficient
number of spatial volumes, with the leading order chiral form of
$f+ge^{-Lm_\pi}/(Lm_\pi)^{3/2}$.  The $L=24a$ values are omitted in
the fit as higher order terms may be required for $Lm_\pi=2.7$. In spite
of the proximity of the $D^{(*)}K$ threshold the volume dependence is small
and for
all decay constants the $L=64a$ data are compatible with the infinite
volume limits. From Table~\ref{tab_1} the largest volume for
$m_\pi=150$~MeV is equivalent in terms of $Lm_\pi$ to the $L=32a$,
$m_\pi=290$~MeV ensemble. For fixed $Lm_\pi$ and to NLO ChPT finite
volume effects are due to one-pion exchange and scale with $g\propto m_\pi^2$,
hence, we estimate these effects to be of the order of
\begin{equation}
\left(f^{290\,\text{MeV}}_{X,L=64a}-f^{290\,\text{MeV}}_{X,L=32a}\right)\times\left(150/290\right)^2
\end{equation}
in the near physical data. In the case of the $D_{s1}$ at the lighter
pion mass one may worry about how to define the decay constants in
view of the possibility of a $p$-wave decay to $D_s\pi\pi$. The
theoretical framework has been
developed in Ref.~\cite{Briceno:2015csa} for two meson channels.
An analogous result does not as yet exist
for the three body problem, however, in view of the narrowness of the
$D_{s1}$ state we would expect such corrections to be very small.

With only one lattice spacing available it is not possible to quantify
the magnitude of discretisation effects. Instead, the 10~MeV
difference between the $a=0.065$~fm result of the ALPHA collaboration
mentioned above and their continuum limit value is taken as an
indication of their possible size. This systematic, along with that
for finite $L$, is included in Table~\ref{tab_6}. We remark that the
shift in the results from a linear chiral extrapolation in $m_\pi^2$
to the physical point is below the statistical standard deviation of
the $m_\pi=150$~MeV results.

Our final results are compared with those of other works in
Table~\ref{tab:comp2}. To our knowledge there is only one previous
lattice study of the decay constants by UKQCD~\cite{Herdoiza:2006qv}
who employ $N_f=2$ non-perturbatively improved clover fermions at a
single coarse lattice spacing of $a=0.10$~fm and a small volume with
$L=1.6$~fm, without consideration of the coupling to the $DK$
threshold. Their values are above ours but in agreement considering
the large uncertainties of their calculation.  Our results are also
somewhat above those derived from the experimental branching ratios of
$B$ decays~(under the assumption of heavy quark symmetry and the
factorisation approximation), while quark model and QCD sum rule
studies give a wide range of values, some of which are consistent with
ours.

In the heavy quark limit the $0^+$ and $1^+$ form a degenerate doublet
with $f_V^{0^+}=f_A^{1^+}$. At the charm quark mass this equality is
violated by 40\%, see Table~\ref{tab:comp2}. As mentioned above the
decay constants are suppressed relative to the corresponding negative
parity ones. This suggests the scalar and axialvector particles are
more spatially extended as might be expected for $P$-wave states but
this is also compatible, for example, with a molecular interpretation.  If
we look to the charmonium sector as an indication of how conventional
$S$- and $P$-wave quark model particles compare, we find the ratio of
decay constants for decay to $\gamma\gamma$ between the
$J^{PC}=0^{-+}$ $\eta_c$ and, the $J^{PC}=0^{++}$ $\chi_{c0}$ is around
0.7.

\section{Conclusions}
\label{conc}

In summary, we have performed a high statistics study of the scalar
and axialvector sectors of the $D_s$ spectrum involving six volumes
comprising linear spatial extents from 1.7~fm up to 4.5~fm and two pion
masses of 290 and 150~MeV for a single lattice spacing
$a=0.07$~fm. The near physical pion mass enables the $DK$ and $D^*K$
thresholds to be realised to within 14~MeV of the QED and isospin
corrected experimental values.  $S$-wave coupling to the threshold is
accounted for in the simulation through the variational approach with
a basis of five quark-antiquark interpolators and a single four quark
interpolator for each channel. The
$D_s^{(*)}\eta$ and $D_s\pi\pi$ thresholds that also exist in the isospin symmetric
limit are not considered.

The four quark operators were found to be essential for reliably
extracting the ground state and first scattering levels in
our setup while in the axialvector channel the third state, identified as
the $D_{s1}(2536)$, could be resolved sufficiently using quark-antiquark
interpolators only. The gap between the first and second scattering levels
is not large for the biggest volumes and the analysis could be 
improved in the future with the inclusion of operators representing
the $D$ and $K$ mesons with opposite momenta.  The quark line diagrams
were evaluated following the stochastic approach of
Refs.~\cite{Aoki:2007rd,Aoki:2011yj,Bali:2015gji}. The limited basis of
interpolators required means this approach is substantially cheaper in
terms of the computer time compared to other methods such as the
distillation technique~\cite{Peardon:2009gh,Morningstar:2011ka} and enables large
volumes and small pion masses to be realised.

The energy spectrum is translated into values for the phase shift
above and below the threshold via L\"uscher's formalism. The data were
consistent with a linear dependence on the energy squared, within the
range $|p^2|\le 300$~GeV$^2$, as expected in the effective range
approximation. The results for the smallest spatial extent of
$L=24a\approx 1.7$~fm lie outside this region and may suffer from
exponentially suppressed finite volume effects which are not included
in the L\"uscher approach or may be in the range where corrections to
linear behaviour are significant.  Our values for the scattering
length, effective range, binding energy and coupling to the threshold
are given in Table~\ref{tab_5}. The scattering lengths are
negative, compatible with the existence of a bound state in each
channel and the infinite volume masses are consistent with the results
from the largest spatial extent of 4.5~fm.  The phase shift was not
evaluated for the $D_{s1}(2536)$ state due to the lack of sensitivity
of the mass to the spatial volume.

A complementary analysis within the chiral unitary approach provided
very similar results for the bound state masses and couplings, see
Table~\ref{tab_7}. One can also access Weinberg's compositeness
probability $1-Z$, which we found to be 1 within errors for both
states. A large value for the latter is often interpreted as
indicating the bound state has a substantial $DK$ component in the
wavefunction.

The final results for the spectrum are compiled in Table~\ref{tab:res}
and displayed in Fig.~\ref{fig:allDs_spectra}. They are comprised of
masses of the $0^+$ and lower $1^+$ state derived from the phase shift
analysis of the $m_\pi=150$~MeV ensemble and the energies of the
negative parity levels and higher $1^+$ state obtained on the largest
spatial volume at this pion mass. Due to the high statistical
precision achieved, significant disagreement is seen with experiment,
in particular for fine structure splittings. The splitting of the
$0^+$ state with the $DK$ threshold is also well below the physical
result, while that for the $1^+$ level is consistent.  These
differences with respect to experiment seem to be predominantly due to
lattice spacing effects, as reasonable agreement is observed for
spin-averaged quantities, for example, for the average threshold
splitting and average $j^P=\frac{1}{2}^+$, $\frac{1}{2}^-$
splitting. Further simulations at finer lattices are required to
remove this source of systematics.

The masses of the scalar and both axialvector particles are sensitive
to the pion mass, suggesting that these may not be conventional quark
model states. A heavier light quark mass leads to more strongly bound
$D_{s0}^*$ and $D_{s1}$ mesons.  Evaluation of the decay constants of
these mesons provides additional inputs to model calculations probing
their internal structure. We find $f_V^{0^+}=114(2)(0)(+5)(10)$~MeV
and $f_A^{1^+}=194(3)(4)(+5)(10)$~MeV, where the errors are due to
statistics, renormalisation, finite volume and lattice spacing
effects. The ratios with the negative parity equivalents are of
similar sizes to those extracted from analyses of non-leptonic $B$
decays to
$D^{(*)}D_{sJ}^{(*)}$~\cite{Datta:2003re,Hwang:2004kga,Cheng:2006dm},
exploiting the factorisation approximation within HQET.  However, our
$f_{V}^{0^+}$ comes out somewhat higher hinting at violations of the
approximations. Finally we also computed the scalar and tensor decay
constants of the $0^+$ and $1^+$ mesons, respectively,
$f_{S}^{0^+}=241(4)(2)(+12)(10)$~MeV and
$f_T^{1^+}=135(2)(2)(+3)(10)$~MeV.  These are not accessible via
leading order Standard Model processes but it would be interesting to see
if any model calculation can reproduce these numbers.

\acknowledgments

We thank Christian Lang and Alberto Martinez Torres for discussions
and Sasa Prelovsek for comments on the manuscript as well as
discussions.  The ensembles were generated primarily on the QPACE
computer~\cite{Baier:2009yq,Nakamura:2011cd}, which was built as part
of the Deutsche Forschungsgemeinschaft SFB/TRR 55 project.  The
authors gratefully acknowledge the Gauss Centre for Supercomputing
e.V. (http://www.gauss-centre.eu) for granting computer time on
SuperMUC at Leibniz Supercomputing Centre (LRZ, http://www.lrz.de) for
this project. Simulations were also performed on the iDataCool cluster
in Regensburg. The BQCD~\cite{Nakamura:2010qh} and
CHROMA~\cite{Edwards:2004sx} software packages were used extensively
along with the locally deflated domain decomposition solver
implementation of openQCD~\cite{luscher3,Luscher:2012av}.

\appendix

\bibliography{threshold}

\end{document}